\documentclass[hidelinks,12pt]{article}
\pdfoutput=1

\usepackage{comment}
\usepackage{epsfig,amssymb,euscript,xspace,xcolor}
\usepackage{amsfonts,amssymb,amsmath,mathtools,empheq,amsthm,graphicx,paralist}
\usepackage{mathrsfs,float}  
\usepackage{pgf,tikz,pgfplots}
\usetikzlibrary{arrows}
\usepackage{tikz,caption,subcaption,marvosym} 
\usetikzlibrary{decorations.markings,arrows,snakes}
\usepackage[belowskip=-15pt,aboveskip=0pt]{caption}
\usepackage{enumitem}

\usepackage[utf8]{inputenc}

\definecolor{lightblue}{rgb}{.1,.4,.5}
\definecolor{brown1}{rgb}{.64,.43,.138}

\usepackage{jheppub}

\def\ZZZ{{\hbox{ Z\kern-1.6mm Z}}}
\def\RRR{{\hbox{ R\kern-2.4mm R}}}
\def\CCC{{\hbox{ C\kern-2.0mm C}}}
\def\zzz{{\hbox{z\kern-1mm z}}}

\newcommand{\qeq}{{\hbox{=\kern-2.3mm ? \kern.5mm }}}
\renewcommand{\qeq}{=}
\usepackage{tikz}

\newcommand{\be}{\begin{equation}}
\newcommand{\ee}{\end{equation}}
\newcommand{\ben}{\begin{eqnarray}\displaystyle}
\newcommand{\een}{\end{eqnarray}}

\def\one{{\hbox{ 1\kern-.8mm l}}}
\def\zero{{\hbox{ 0\kern-1.5mm 0}}}

\newcommand{\bea}[1]{\begin{eqnarray}\label{#1} }
\newcommand{\eea}{\end{eqnarray}}

\usepackage{bm}

\def\bea{\begin{eqnarray}}
\def\eea{\end{eqnarray}}
\def\be{\begin{equation}}
\def\ee{\end{equation}}

\definecolor{wvvxds}{rgb}{0.396078431372549,0.3411764705882353,0.8235294117647058}
\definecolor{dbwrru}{rgb}{0.8588235294117647,0.3803921568627451,0.0784313725490196}
\definecolor{dtsfsf}{rgb}{0.8274509803921568,0.1843137254901961,0.1843137254901961}
\definecolor{wrwrwr}{rgb}{0.3803921568627451,0.3803921568627451,0.3803921568627451}
\definecolor{cqcqcq}{rgb}{0.7529411764705882,0.7529411764705882,0.7529411764705882}
\definecolor{rvwvcq}{rgb}{0.08235294117647059,0.396078431372549,0.7529411764705882}


\date{\today}
\title{Stringy canonical forms and binary geometries from associahedra, cyclohedra and generalized permutohedra}

\begin{document}

\author[a,b,c,d]{Song He}
\author[a,d]{,\,Zhenjie Li}
\author[a,e,f]{,\,Prashanth Raman}
\author[a,d]{,\,Chi Zhang}
\affiliation[a]{CAS Key Laboratory of Theoretical Physics, Institute of Theoretical Physics, Chinese Academy of Sciences, Beijing 100190, China}
\affiliation[b]{
School of Fundamental Physics and Mathematical Sciences, Hangzhou Institute for Advanced Study, UCAS, Hangzhou 310024, China}
\affiliation[c]{ICTP-AP
International Centre for Theoretical Physics Asia-Pacific, Beijing/Hangzhou, China}
\affiliation[d]{School of Physical Sciences, University of Chinese Academy of Sciences, No.19A Yuquan Road, Beijing 100049, China}
\affiliation[e]{Institute of Mathematical Sciences, Taramani, Chennai 600 113, India}
\affiliation[f]{Homi Bhabha National Institute, Anushakti Nagar, Mumbai 400085, India}
\emailAdd{songhe@itp.ac.cn}
\emailAdd{lizhenjie@itp.ac.cn}
\emailAdd{prashanthr@imsc.res.in}
\emailAdd{zhangchi@itp.ac.cn}
\date{\today}
\abstract{Stringy canonical forms are a class of integrals that provide $\alpha'$-deformations of the canonical form of any polytopes. For generalized associahedra of finite-type cluster algebras, there exist completely rigid stringy integrals, whose configuration spaces are the so-called binary geometries, and for classical types are associated with (generalized) scattering of particles and strings. In this paper, we propose a large class of rigid stringy canonical forms for another class of polytopes, generalized permutohedra, which also include associahedra and cyclohedra as special cases (type $A_n$ and $B_n$ generalized associahedra). Remarkably, we find that the configuration spaces of such integrals are also binary geometries, which were suspected to exist for generalized associahedra only. For any generalized permutohedron that can be written as Minkowski sum of coordinate simplices, we show that its rigid stringy integral factorizes into products of lower integrals for massless poles at finite $\alpha'$, and the configuration space is binary although the $u$ equations take a more general form than those ``perfect'' ones for cluster cases. Moreover, we provide an infinite class of examples obtained by degenerations of type $A_n$ and $B_n$ integrals, which have perfect $u$ equations as well. Our results provide yet another family of generalizations of the usual string integral and moduli space, whose physical interpretations remain to be explored. }

\maketitle

\section{Introduction}
In \cite{Arkani-Hamed:2019mrd} the notion of {\it stringy canonical forms} was introduced which provides vast generalizations of tree-level string amplitudes as integrals over the moduli space ${\cal M}_{0, n}$~\footnote{The study of these types of integrals perhaps dates back to Euler, and more recently they are studied in various contexts, see \textit{e.g.} \cite{Mellin, Eulermellin, Panzer:2019yxl, Brown:2009qja, aomoto2011theory, Mizera:2017rqa, Mastrolia:2018uzb, Mizera:2019vvs,Brown:2018omk}. 
}. Various nice properties of string amplitudes and their field-theory limits, such as factorizations for $\alpha' \to 0$ limit and finite $\alpha'$, channel duality and exponential softness at UV, scattering equations, {\it etc.} all naturally extend to stringy canonical forms. To briefly summarize the idea, let's begin with the disk integral for the open-string amplitudes with ordering $(1,2,\dots, n)$:
\begin{equation}\label{stringint}
{\cal I}^{\rm open-string}_{1,2,\dots,n} (\{s\}):=\int_{\mathcal M_{0,n}^+}
\frac{\mathrm d^{n}z}{\mathrm{SL}(2,\mathbb{R})} \prod_{i=1}^{n}\frac{1}{z_{i}-z_{i{+}1}}
\prod_{i<j}(z_j{-}z_i)^{\alpha' s_{ij}},
\end{equation}
where $\mathcal M_{0,n}^+$ is the positive part of moduli space for the ordering, and $\mathsf{PT}_n:=\frac{\mathrm d^{n}z}{\mathrm{SL}(2,\mathbb{R})} \prod_{i=1}^{n}\frac{1}{z_{i}-z_{i+1}}$ its canonical form~\cite{Arkani-Hamed:2017mur}.  
With a positive parametrization $\{x_1,\dots,x_{n-3}\}$ of $\mathcal M_{0,n}^+$, 
this canonical form becomes $\prod_{i=1}^{n-3}\mathrm d \log x_i$, and the Koba-Nielsen factor $\prod_{i<j}(z_j-z_i)^{\alpha' s_{i j}}$ becomes a product of powers of polynomials with 
non-negative coefficients~\cite{Arkani-Hamed:2019mrd}: $\prod_i x_i^{\alpha' X_i} \prod_I p_I(x)^{-\alpha' c_I}$. Note that we have factored out $n{-}3$ monomials, $x_i$, and left with $n(n{-}3)/2-(n{-}3)$ positive polynomials $p_I(x)$, with exponents $X_i$'s and $c_I$'s given by linear combinations of the Mandelstam variables $s_{ij}$. 

\paragraph{Stringy canonical forms} Motivated by this rewriting of string integral~\eqref{stringint}, we define a general stringy integral as a $d$-dimensional integral~\cite{Arkani-Hamed:2019mrd}
\begin{equation}\label{stringycanonicalform}
	\mathcal I (\mathbf X,\{c\})=
	\int_{\mathbb R_+^d}\prod_{i=1}^d \frac{\mathrm d x_i}{x_i}x_i^{\alpha'X_i}
	\prod_{I=1}^m p_I(x)^{-\alpha' c_I},
\end{equation}
where we allow $m$ arbitrary Laurent polynomials with non-negative coefficients, $p_I(x)$, and without loss of generality we assume exponents $X_i>0$ and $c_I>0$~\footnote{These are ``positive'' stringy canonical forms, which are integrals over $\mathbb R_+^d$. We can also consider ``complex'' cases with the integrand mod-squared and integrated over $\mathbb C^d$, which are generalizations of closed-string amplitudes~\cite{Arkani-Hamed:2019mrd}.}. 

Any integral of the form \eqref{stringycanonicalform} inherits various remarkable properties of string amplitudes \eqref{stringint}, which all depends on the existence of a polytope $\mathcal P:=\sum_I c_I N[p_I]$; here $N[p_I]$ is the Newton polytope of the polynomial $p_I$, and we take the (weighted) Minkowski sum~\cite{Arkani-Hamed:2019mrd}. First of all, one can show that the integral \eqref{stringycanonicalform} is convergent if and only if $\mathbf X=(X_1,\dots,X_d)$ is in the interior of the polytope $\mathcal P$ (we assume that ${\cal P}$ is $d$ dimensional). The leading order of the integral in its $\alpha'$ expansion, dressed with $\mathrm{d}^d \mathbf X$, is given by the canonical form of the polytope, ${\Omega}(\cal P)$, thus it can be viewed as an $\alpha'$-deformation of $\Omega(\cal P)$ (hence the name ``stringy canonical form''). Generically the poles of $\Omega(\cal P)$ are given by equations of
facets for $\mathcal P$, which are also poles of the integral at finite $\alpha'$, and the residue on any such pole is given by a stringy canonical form defined for that facet (see~\cite{Arkani-Hamed:2019mrd} for details). 

For the string integral \eqref{stringint}, ${\cal P}$ is nothing but the ABHY associahedron~\cite{Arkani-Hamed:2017mur}, whose canonical form gives bi-adjoint $\phi^3$ amplitude as the $\alpha'\to 0$ limit of \eqref{stringint}~\cite{Cachazo:2013iea}. The poles are the planar variables $X_{i,j}$, on which \eqref{stringint} factorizes into lower-point integrals even for finite $\alpha'$. Moreover, the saddle point equations as $\alpha' \to \infty$ can be written as $X_i=\sum_I c_I \partial \log p_I/\partial \log x_i$ for $i=1,\dots, d$; they provide a diffeomorphism from the integration domain to ${\cal P}$, thus $\Omega(\mathcal P)$ can be computed as a pushforward of $\prod_i \mathrm{d}\log x_i$ via summing over the saddle points~\cite{Arkani-Hamed:2017tmz}. In particular, for the string integral \eqref{stringint}, the scattering equations provide a map from ${\cal M}_{0,n}^+$ to ABHY associahedron, and the pushforward is equivalent to the CHY formula for $\phi^3$ amplitude~\cite{Cachazo:2013hca,Cachazo:2013iea}. 

\paragraph{Configuration spaces} 
One way to describe the combinatorics of the polytope ${\cal P}$ of the stringy integral \eqref{stringycanonicalform} is through the {\it configuration space}, \textit{e.g.}
as a blow-up of the integration domain, $\mathbf x:=(x_1,\dots,x_d)$-space $\mathbb R_+^d$. 
However, it is usually difficult to use $\mathbf x$ to describe the polytope $\mathcal P$, 
even for the string integrals \eqref{stringint}. For example, a facet usually corresponds to a complicated limit process with $x\to 0$ or $\infty$.
As proposed in~\cite{Arkani-Hamed:2019plo}, it's more elegant to realize ${\cal P}$ by introducing a set of constrained \textit{$u$ variables} $\{u_\alpha\}$ which satisfy a set of polynomial equations called \textit{$u$ equations}. The configuration space is the variety of $u$ variables defined by $u$ equations, where the polytope ${\cal P}$ can be realized combinatorially by $\{u_\alpha \geq 0\}$ such that each facet $F_\alpha$ of ${\cal P}$ is simply given by an equation $u_\alpha=0$ and its interior is described by $\{u_\alpha > 0\}$.
To define the corresponding $u$ variables of the stringy canonical form \eqref{stringycanonicalform} from $\mathbf x$-space $\mathbb R_+^d$, we first construct the so-called \textit{big polyhedron} in the $\mathbf S:=(X_1,\dots,X_d,-c_1,\dots,-c_m)$-space $\mathbb R^{d{+}m}$.  

The polytope ${\cal P}$ in the space of $\mathbf X:=(X_1,\dots,X_d)$ is bounded by inequalities $W_a^JS_J\geq 0$, where $W_a^J$ is determined by the facet $F_a$ of $\mathcal P$. These inequalities also cut out a polyhedron in the $(d+m)$-dimensional space of $\mathbf S$, which is defined to be the big polyhedron of $\mathcal P$, 
and the integral $\mathcal I(\mathbf S)$ converges for each point $\mathbf S$ 
inside this polyhedron. Dually, we can describe the big polyhedron by its vertices. Suppose 
$\{V_J^\alpha\}_{1\leq \alpha \leq v}$ are its non-zero vertices ($v$ denotes the number of vertices), a point $\mathbf S$ inside 
the polyhedron can be written as $S_J = F_\alpha V^\alpha_J$ with non-negative coefficients 
$\{F_\alpha\}_{1\leq \alpha \leq v}$.  Therefore, the integral $\mathcal I(\mathbf S)$
can be seen as a function of ``dual coordinates'' $\{F_\alpha\}$ which encourages us to 
rewrite the integral to make it manifest by (here we collectively denote $p_J=(x_1, \cdots, x_d, p_1, \cdots, p_m)$)~\footnote{Note that one can indeed treat the monomials $x_i^{\alpha' X_i}$ on the equal footing as polynomials, since Minkowski sum w.r.t. them amounts to merely translating the polytope $c_I N[p_I]$ by $-X_i$ in the $i$-th direction~\cite{Arkani-Hamed:2019mrd}.}
\begin{equation}
	\mathcal I(\mathbf F)=
	\int_{\mathbb R_+^d} \prod_{i=1}^d \frac{d x_i}{x_i}~
	\prod_{\alpha}u_\alpha^{\alpha' F_\alpha}
	=
	\int_{\mathbb R_+^d}\prod_{i=1}^d \frac{d x_i}{x_i}~
	\prod_{J}^{d{+}m} p_J^{\alpha' S_J},
\end{equation}
where we introduce a new set of variables $u_\alpha = \prod_J p_J^{V_J^\alpha}$ for 
$\alpha=1,\dots, v$. Note that there's no unique way to write $S_J=F_\alpha V_J^\alpha$ if the big polyhedron is not a simplex, \textit{i.e.} $v>d+m$, and in this case,
$u$ variables $\{u_\alpha \}$ are not multiplicatively independent. Therefore, it is an important special case when the big polyhedron is a simplex, \textit{i.e.} $N=v=d+m$, where $N$ is 
the number of facets of the big polyhedron or the original polytope $\mathcal P$,
and inequalities $F_\alpha=S_J(V^{-1})^J_\alpha \geq 0$ are exactly the inequalities $W_a^J S_J\geq 0$ up to rescaling $F_\alpha \mapsto t F_\alpha$ for positive factors $t$, because
there's no nontrivial linear isometry from the simplex to itself besides permutations of vertices. Therefore, we identity the label of facets $\alpha$ with the label of 
vertices $J$ and get $V=W^{-1}$. Furthermore, every facet of the original polytope $\mathcal P$ can be associated with a single $u_\alpha$ going to zero. 

\paragraph{Cluster stringy integrals and binary geometries} A class of very special integrals belong to this case~\cite{Arkani-Hamed:2019plo}: stringy canonical forms for generalized associahedra of finite-type cluster algebras, namely $A_n$, $B_n$, $C_n$, $D_n$, and exceptional cases ($E_6$, $E_7$, $E_8$, $F_4$, $G_2$). These ``cluster string integrals''~\cite{Arkani-Hamed:2019plo}, and the associated ``cluster configuration spaces''~\cite{Arkani-Hamed:2019plo,Arkani-Hamed:2020tuz}, turn out to be much more special and rigid extensions of string amplitudes \eqref{stringint} and the moduli space ${\cal M}_{0,n}$, which correspond to the case of type $A_{n{-}3}$. The leading order of such an integral is given by the canonical function of the corresponding generalized associahedra, whose combinatorics is attached to the Dynkin diagram (each facet is given by removing a node), {\it e.g.} $D_n \to D_m \times A_{n{-}1{-}m}$ or $A_{n{-}1}$~\cite{Arkani-Hamed:2019plo}. Even at finite $\alpha'$, it has the remarkable factorization property tied to the Dynkin diagram. Moreover, for type $B/C$ (known as ``cyclohedron'') and $D$, the leading orders are associated with tadpole emission and one-loop bi-adjoint $\phi^3$ amplitudes, respectively~\cite{Salvatori:2018aha,Arkani-Hamed:2019vag}.

These nice properties become manifest in the configuration space: not only the big polyhedron is a simplex, the $u$ equations are of very special form:
\begin{equation}\label{perfectu}
	1-u_\alpha=\prod_{\beta} u_\beta^{\beta||\alpha},
\end{equation}
where the product is over all $N$ variables ($N$ facets), and the integer $\beta || \alpha\geq 0$ is the so-called compatibility degree defined in the corresponding cluster algebra~\cite{Zelevinskysys}. Varieties defined by $u$ equations associated with finite type clusters  are called cluster configuration spaces.
We see that when $u_\alpha$ (for a facet $\alpha$) goes to $0$, it forces all the $u_\beta$ with $\beta || \alpha>0$ to go to $1$ (for facets $\beta$ are exactly those {\it incompatible} with $\alpha$); the remaining $u$'s that are compatible with $u_\alpha$ satisfy $u$ equations for the facet, which factorizes as removing a node of the Dynkin diagram. Therefore, we say that the configuration space is a {\it binary geometry}. Note that this is true also for the complex case, and if we further require all $u$ variables to be positive, we have $0<u<1$ which gives the positive configuration space~\cite{Arkani-Hamed:2020cig}. The binary property is what guarantees the stringy integral to factorize even at finite $\alpha'$: on the pole $X_\alpha \to 0$, $u_\alpha\to 0$ forces all incompatible $u_\beta\to 1$ which are decoupled from $\prod u^{\alpha' X}$, and the integral becomes the factorized one for the facet $\alpha$. 

For example, the string integral \eqref{stringint} corresponds to type $A_{n{-}3}$, 
and the $N=n(n{-}3)/2$ $u$ variables of this configuration space are~\cite{Arkani-Hamed:2020cig}
\[
	u_{i j}=\frac{(z_{i-1}-z_j)(z_i-z_{j-1})}{(z_i-z_j)(z_{i-1}-z_{j-1})},
\]
and they satisfy the $u$ equations (see {\it e.g.}~\cite{Brown:2009qja})
\begin{equation}\label{Aueq}
	1-u_{ij}=\prod_{(k,l)\not\sim (i,j)} u_{kl},
\end{equation}
where $(i,j)\not\sim (k,l)$ means that the diagonals $(i,j)$ and $(k,l)$ of $n$-gon are crossed, thus the two facets are incompatible (note here compatibility degree is just $0$ or $1$).  For $u_{i,j}\to 0$, all incompatible $u_{k,l} \to 1$, and the remaining $u$'s fall into those of the two polygons divided by $(i,j)$, with their own $u$ equations. The Koba-Nielsen factor becomes $\prod_{i,j} u_{i,j}^{\alpha' X_{i,j}}$ where $X_{i,j}$ are planar variables for the facets of ABHY associahedron, which makes the factorization of string amplitude at finite $\alpha'$ manifest. All these generalize to integrals and configuration spaces for finite type cluster algebras. 

\subsection{Summary of main results}

It is a natural question to ask if binary geometries, whose stringy integrals factorize at any finite $\alpha'$, are extremely special and can only be products of generalized associahedra of finite type. In this paper we answer the question in the negative by providing infinitely many new examples of binary geometries. Before proceeding, we remark that $u$ equations of the form \eqref{perfectu} (which we refer to as ``perfect'' $u$ equations) imply binary geometries, but the converse is not true. Broadly speaking, for a $n$-dimensional polytope ${\cal P}$ where each facet is associated with a $u$ variable, if for any $u_\alpha\to 0$, we have for all incompatible facets $\beta$, $u_\beta \to 1$, then such an $n$-dimensional variety (defined by the $u$ equations) is a binary geometry. From this definition, we see that the simplest example of binary geometry is a simplex, where all facets are compatible with each other. One can realize the configurations space by the following $u$ equations for a $n$-dimensional simplex with facets labelled by $\alpha=0,1,\dots, n$:
\[
1-u_\alpha=\sum_{\beta \neq \alpha} u_\beta\,.
\] 
As $u_\alpha\to 0$, there is no $u\to 1$, and all $n$ remaining $u$ variables satisfy the same equations for that facet, which is an $(n{-}1)$-dimensional simplex. Although this example is trivial, it does show that we do not necessarily need perfect $u$ equations to have binary geometries. Generically,  a $n$-dimensional polytope ${\cal P}$ with $N$ facets have an $n$-dimensional configuration space defined by $u$ equations
\be\label{genueq}
1-u_\alpha=p_\alpha(\{u\}) \prod_\beta u_\beta^{\beta||\alpha} 
\ee
where $\beta ||\alpha \geq 0$ are integers which vanishes for $\beta$ compatible with $\alpha$, and $p_\alpha(\{u\})$ is an arbitrary polynomial which must equal to unity on the facet $\alpha$, $p_\alpha (\{u\})|_{u_\alpha=0}=1$. When $u_\alpha \to 0$, since it appears on the RHS of the equation with $1-u_\beta$ for any incompatible $\beta$, we have $u_\beta\to 1$ for any polynomial $p_\beta (\{u\})$. However, in the equation with $1-u_\alpha$, since all the incompatible $u_\beta \to 1$ we find $p_\alpha (\{u\})|_{u_\alpha=0}=1$ (where we evaluate the polynomial at $u_\alpha=0$ and all incompatible $u_\beta=1$). For cluster configuration spaces, we have perfect $u$ equations with $p_\alpha=1$ for all $\alpha$.

As the main result of the paper, we present infinitely many new examples of binary geometries which have $u$ equations of the form \eqref{genueq}, and (also infinitely) many of them with perfect $u$ equations. The key idea to consider a class of stringy integrals for {\it generalized permutohedra} that can be realized as the Minkowski sum of coordinate simplices is as follows (more details will be given later). Given the label set $[0,n]=\{0,1,\dots, n\}$, we define a {\it building set} $\mathbf B$ as a collection of subsets $I \subset [0,n]$ that satisfy (1). if $I, J \in {\mathbf B}$ and $I\cap J\neq \varnothing$, then $I\cup J \in {\mathbf B}$, and (2). it contains all {\it singletons} $\{i\}$. For each subset $I$ we have a simplex $\Delta_I$ as the convex hull of unit vectors ${\bf e}_i$ with $i\in I$, and the generalized permutohedron ${\mathscr P}(\mathbf B)$ is given by (weighted) Minkowski sum of $\Delta_I$ (defined on a $n$-dimensional hyperplane of the $(n{+}1)$-dimensional space)~\cite{Postnikov:2005}.  It is very natural to define stringy canonical form for such a building set $\mathbf B$: for each subset $I$, we have the simplest {\it linear} polynomial $x_I:=\sum_{i\in I} x_i$ (for singletons, $x_{\{i\}}=x_i$ are the coordinates), and
\begin{equation}\label{intgenperm}
{\cal I}_{\mathbf B}(\{S\}):=(\alpha')^n~\int_{\mathbb R_+^n}\frac{1}{{\rm vol}~GL(1)} \prod_{i=0}^n \frac{d x_i}{x_i} 
\prod_{I \in {\mathbf B}} x_I^{\alpha' S_I}\,,
\end{equation}
where the integration domain is defined projectively, {\it i.e.} up to a $GL(1)$ redundancy thus the exponents $S_I$ satisfy $\sum_I S_I=0$ (one can fix the gauge by choosing {\it e.g.} $x_0=1$, then $S_0$ drops out and the remaining $S_I$ are independent). By construction, ${\cal I}_{\mathbf B}$ is a stringy canonical form for the generalized permutohedron $\sum_I S_I {\bf N} [x_I]=\sum_I S_I \Delta_I={\mathscr P}(\mathbf B)$, and the leading order is given by the volume of its dual. Note that one can split the product as $\prod_I x_I^{\alpha' S_I}=\prod_{i=1}^n x_i^{\alpha' X_i} \prod_I^m x_I^{-\alpha' C_I}$ with $X_i:=S_{\{i\}}$ for singletons, $-C_I=S_I$ for non-singleton $I$'s (we denote their total number as $m$), and it suffices to take the Minkowski sum w.r.t. the non-singleton $I$'s, since the singleton ones are merely translating the polytope. Let's first present the simplest example with building set ${\mathbf B}$ given by all the singletons $\{i\}$ and $[0,n]$. In this case, ${\mathscr P}(\mathbb B)=c {\bf N} [\sum_{i=0}^n x_i]$ is the $n$-dimensional simplex (we denote $c:=C_{[0, n]}$), given by $X_i \geq 0$ with one (ABHY) constraint $\sum_{i=1}^n X_i=c$. It is straightforward to see that the $u$ variables are $u_i=\frac{x_i}{x_{[0,n]}}=\frac{x_i}{\sum_i x_i}$ and they satisfy $\sum_{i=0}^d u_i=1$ as expected. In this case, the integral is trivial to perform, which yields a generalization of beta function to higher dimensions:
\[
{\cal I}^{\rm simplex}_n=\frac{\prod_{i=1}^n \Gamma(\alpha' X_i)\Gamma(\alpha'(c- \sum_i X_i))}{\Gamma(\alpha'c)}:=B\left(\alpha' X_1, \dots, \alpha' X_n,\alpha'\left(c- \sum_i X_i\right)\right)\,.
\]

The most important three examples of ${\mathscr P} (\mathbf B)$ and ${\cal I}_{\mathbf B}$, which will be studied in detail throughout the paper, are summarized here:
\begin{itemize}
\item Consider $\mathbf B$ as the collection of all $n(n{+}3)/2+1$ consecutive intervals of $[0,n]$, ${\mathscr P}(\mathbf B) \equiv {\mathscr A}_n$ is the $n$-dimensional associahedron, \eqref{intgenperm} is nothing but $(n{+}3)$-pt string integral \eqref{stringint}; {\it e.g.} for $n=2$, ${\mathbf B}=\{\{i\}, \{0,1\}, \{1,2\}, \{0,1,2\}\}$, and we have a pentagon ${\mathscr A}_2$.
\item Consider $\mathbf B$ as the collection of all $n(n{+}1)+1$ cyclic intervals of $[0,n]$ which can wrap around $n$, ${\mathscr P} (\mathbf B) \equiv {\mathscr B}_n$ is the $n$-dimensional cyclohedron, and in sec. 2 we will show that non-trivially \eqref{intgenperm} is the cluster stringy integral for type $B_n$ (not that for type $C_n$); {\it e.g.} $n=2$, ${\mathbf B}=\{\{i\}, \{0,1\}, \{1,2\}, \{2,0\}, \{0,1,2\}\}$, and we have a hexagon ${\mathscr B}_2$.
\item Consider $\mathbf B$ as the collection of all $2^{n{+}1}-1$ nonempty subsets of $[0,n]$, ${\mathscr P}(\mathbf B) \equiv {\mathscr P}_n$ is the $n$-dimensional permutohedron (for $n=2$, it coincides with ${\mathscr B}_2$). All generalized permutohedra (and their stringy integrals) can be obtained from (that of) ${\mathscr P}_n$. 
\end{itemize}

Although associahedra and cyclohedra are special cases of generalized permutohedra, it is interesting that \eqref{intgenperm} in these cases turn out to be exactly cluster stringy integrals of type $A$ and $B$! It is not completely surprising since these are the only two cases of cluster configuration spaces that can be identified with hyperplane arrangement~\cite{Arkani-Hamed:2020tuz}, which is consistent with the fact that here we only have linear polynomials $x_I$. In sec. 2, we will go further and discover more binary geometries with perfect $u$ equations, which can be obtained by ``degenerating'' type $A$ and $B$ integrals by requiring certain $S_I=0$ (for example, type $A$ integral is in fact a degeneration of the type $B$ integral, which was not obvious before). Some of these cases are products of lower-dimensional $A$'s and $B$'s, but we will see entirely new binary geometries with perfect $u$ equations.

Moreover, a remarkable statement we will make in sec. 3 is that, the configuration space for any ${\cal I}_{\mathbf B}$ is always binary, which means that any $u_\alpha \to 0$ forces all the incompatible ones $u_\beta\to 1$. As we will see explicitly with ${\mathscr P}_n$, its $u$ equations indeed take the form of \eqref{genueq}, which is no longer perfect for $n>2$, but the space is binary and the integral factorizes as a product of lower-dimensional permutohedron integrals at finite $\alpha'$! Note that an important feature of any generalized permutohedron is that its facets are labelled by all subsets $I$ in ${\mathbf B}$ except for the complete set $[0,n]$, and on each facet it factorizes as a product of two generalized permutohedra. The total number of facets $N$ equals $|{\mathbf B}|-1=n+m$, thus the big polyhedron is indeed a simplex as expected. In sec. 3, we will present the formula for $u$ variables for any generalized permutohedron (which is equivalent to the formula for its ABHY-like realization), from which the binary and factorization properties follow.

\section{Configuration spaces with perfect $u$ equations from degenerating ${\mathscr A_n}$ and ${\mathscr B_n}$}
In this section we shall consider some degenerations of $n$-dimensional associahedron ${\mathscr A_n}$ and $n$-dimensional cyclohedron ${\mathscr B_n}$ and show that there are an infinite class of examples of binary geometries with perfect $u$ equations. To do this we shall use the fact that both ${\mathscr A_n}$ and ${\mathscr B_n}$ can be realised as a Minkowski sum of coordinate simplices. 

\subsection{${\mathscr A_n}$ and ${\mathscr B_n}$ as generalized permutohedra} 
\paragraph {Associahedron $\mathscr A_n$} The building set of $\mathscr A_n$ is $\mathbf B= \{[i,j]\,:\, 0\leq i\leq j\leq n\}$.
The stringy canonical form of this building set $\mathcal I_{\mathbf B}(\{S\})$
is related with the original string integral eq.\eqref{stringint} by 
\[
	z_j-z_i=x_{[i,j-1]},
\]
so Mandelstam variables are related to $S$ by $s_{ij}=S_{[i,j-1]}$.
The facets of $\mathscr A_n$ are given by
\[
	F_{[i,j]}=\sum_{[k,l]\subset [i,j]}S_{[k,l]},
\]
and their solution is (a rewriting of ABHY conditions in our notation):
\begin{equation}\label{Ansf}
S_{[i,j]}=F_{[i,j]}+F_{[i+1,j-1]}-F_{[i,j-1]}-F_{[i+1,j]},
\end{equation}
where $F_{[k,l]}:=0$ when $k>l$ or $[k,l]=[0,n]$ and the usual planar $X$ variables $X_{ij}
$ is $F_{[i,j-2]}$. Using this solution to rewrite the integrand of the stringy canonical form $\mathcal I_{\mathbf B}(\{S\})$
\begin{align*}
\prod_{0\leq i\leq j\leq n}x_{[i,j]}^{S_{[i,j]}}
&=\prod_{0\leq i\leq j\leq n}x_{[i,j]}^{F_{[i,j]}}x_{[i,j]}^{F_{[i+1,j-1]}}\bigg /
x_{[i,j]}^{F_{[i,j-1]}}x_{[i,j]}^{F_{[i+1,j]}}
\end{align*}
and we get that the $u$ variables easily:
\begin{equation}
 u_{i,j+2}:=u_{[i,j]}=\frac{x_{[i,j]}x_{[i-1,j+1]}}{x_{[i,j+1]}x_{[i-1,j]}} \quad \text{for $0\leq i\leq j\leq n$,}   \label{uvarforAn}
\end{equation}
with $x_{[i,n+1]}:=1$ and $x_{[-1,j]}:=1$, and it is straightforward to see that they satisfy $u$ equations \eqref{Aueq}.

The ABHY construction of $\mathscr A_n$ is given by 
\[
	S_{[i,j]}=-C_{[i,j]} \,\,\text{ for all $i<j$,}\quad
	F_{[i,j]}\geq 0 \,\,\text{ for all $i\leq j$},
\]
where $C$'s are positive numbers. As first introduced in~\cite{Arkani-Hamed:2019vag,Bazier-Matte:2018rat}, it is convenient and illuminating to represent these conditions in a mesh diagram of $(1+1)$-D lattice. For example, we can have the following mesh for ABHY realization of $\mathscr A_3$:  
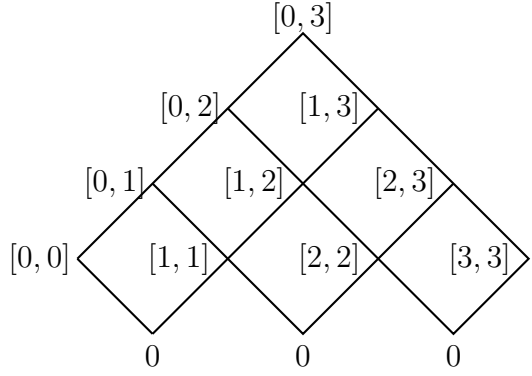
\begin{center}
\begin{tikzpicture}
	\draw[thick] (-2,-1) -- (1,2) -- (4,-1) -- (3,-2) node[below]{0} -- (0,1);
	\draw[thick] (-1,-2) node[below]{0} -- (2,1);
	\draw[thick] (-2,-1) -- (-1,-2);
	\draw[thick] (-1,0) -- (1,-2) node[below]{0}-- (3,0);
	\node at (-2.5,-1) {$[0,0]$};
	\node at (-0.65,-1) {$[1,1]$};
	\node at (1.35,-1) {$[2,2]$};
	\node at (3.35,-1) {$[3,3]$};
	\node at (-1.5,0) {$[0,1]$};
	\node at (0.35,0) {$[1,2]$};
	\node at (2.35,0) {$[2,3]$};
	\node at (-0.5,1) {$[0,2]$};
	\node at (1.35,1) {$[1,3]$};
	\node at (1,2.2) {$[0,3]$};
\end{tikzpicture}
\end{center}
where each diamond is labelled by $[i,j]$ of the top corner~\footnote{
It looks different with the usual mesh of $\mathscr A_n$~\cite{Arkani-Hamed:2019vag} where label of $c$ follows the left corner
in each diamond. This is because the mismatch of the correspondence of labels 
$X_{ij}=F_{[i,j-2]}$ and $s_{ij}=S_{[i,j-1]}$.
}, and it gives
\begin{equation}\label{CinFA}
	C_{[i,j]}=F_{[i,j-l]}+F_{[i+1,j]}-F_{[i,j]}-F_{[i+1,j-1]}\,.
\end{equation}
For each causal diamond, we have the Gauss law:
\[
	\sum_{i\leq a<j\atop{k<b\leq l}} c_{[a,b]} = 
	F_{[i,k]}+F_{[j,l]}-F_{[i,l]}-F_{[j,k]}\,,
\]
where $F_{[a,b]}:=0$ when $a>b$ or $[a,b]=[0,n]$

\paragraph{Cyclohedron $\mathscr B_n$} The building set of $\mathscr B_n$ is $\mathbf B=\{[i,i+k]\,:\,0\leq i, k\leq n\}$ where the labels are understood to be mod $n+1$. The facets are given by 
\[
F_{[i,j]}=\sum_{[k,l]\subset [i,j]}S_{[k,l]},
\]
and their solution is 
\[
	S_{[i,i]}=F_{[i,i]},\quad S_{[i,j]}=F_{[i,j]}+F_{[i+1,j-1]}-F_{[i,j-1]}-F_{[i+1,j]}\quad \text{for $i\neq j$ and $[i,j]\neq [0,n]$},
\]
and
\[
	S_{[0,n]}=F_{[0,n]}-\sum_{i=0}^n F_{[i,i-2]} +\sum_{i=0}^n F_{[i,i-3]},
\]
where $F_{[i+1,i]}=F_{[0,n]}=0$. Therefore, the $u$ variables are
\begin{equation}
   u_i:=u_{[i,i-2]}=\frac{x_{[i,i-2]}}{x_{[0,n]}},\quad 
u_{i,j+2}:=u_{[i,j]}=\frac{x_{[i,j]}x_{[i-1,j+1]}}{x_{[i,j+1]}x_{[i-1,j]}}
\quad \text{for $i-j\neq 2$}.  \label{uvarforBn}
\end{equation}
It is straightforward to work out the $u$ equations, which can be put in a compact form~\cite{Arkani-Hamed:2019plo}
\begin{equation}\label{Bueq}
1-u_i=U_{[i+1,i-1]},\quad 
1-u_{ij}=U_{[i+1,j],[j+1,i-1]}U_{[j+1,i-1],[i,j-1]}U^2_{[j+1,i-1]},
\end{equation}
where we introduce the notation 
\[
U_{A,B}=\prod_{a\in A,b\in B}u_{ab},\quad 
U_A=\prod_{a\in A}u_a\prod_{a<b\in A}u_{ab}.
\]

Since the building set of $\mathscr A_n$ is contained in the building set of $\mathscr B_n$, $\mathscr A_n$ is a degeneration of $\mathscr B_n$, which was not obvious in the previous construction of $\mathscr B_n$.  

The ABHY construction of $\mathscr B_n$ is similar: $F_{[i,j]}\geq 0$ for all $[i,j]\in\mathbf B$ and for $i\neq j$
\begin{equation}\label{abhyB}
F_{[i,j-1]}+F_{[i+1,j]}-F_{[i,j]}-F_{[i+1,j-1]}
	=C_{[i,j]}\,\,\text{for $[i,j]\neq [0,n]$}, \quad C_{[0,n]}=\sum_{i=0}^n F_{[i,i-2]}-\sum_{i=0}^n F_{[i,i-3]},
\end{equation}
where $F_{[i+1,i]}=0$ and the $C$'s are all positive constants. 

Like $A_n$, $B_n$ type cluster stringy canonical form also comes from a hyperplane 
arrangement, the Shi arrangement~\cite{shi1986kazhdan}. The Shi arrangement contains $n+1$ punctures 
$\{z_i\}_{i=1,\dots,n+1}$ on the real line with the freedom of global transformation 
$z_i\to z_i+a$ which can be used to fix $z_{n+1}=0$. The Shi arrangement is 
given by the following hyperplanes
\[
	z_i-z_j=0\quad \text{and}\quad z_i-z_j=1\quad \text{for $1\leq i<j\leq n+1$},
\]
and its stringy integral is 
\begin{equation}
\mathcal I_n = \int_{1>z_1>z_2>\cdots >z_n>0}
\frac{\mathrm dz_1\cdots \mathrm dz_{n}}{(1-z_1)(z_1-z_2)\cdots (z_{n}-0)}
\prod_{1\leq i<j \leq n+1}(z_i-z_j)^{s_{ij}}(1-z_i+z_j)^{t_{ij}},
\end{equation}
where the positive region is given by $0<z_i-z_j<1$ for $1\leq i<j \leq n+1$.
We can also write the $u$ variables in terms of the $z$'s:
\begin{align*}
    &u_1=\tilde z_1-z_2,\quad \dots,\quad u_{n}=\tilde z_{n}-z_{n+1},
    \quad u_{n+1}=z_{n+1}-z_1,\\
    &u_{ji}=\frac{(z_{j+1}-z_{i})(z_{i+1}-z_j)}{(z_{j+1}-z_{i+1})(z_i-z_j)}\quad \text{for $i<j$},\\
    &u_{ij}=\frac{(\tilde z_{j+1}-z_{i})(\tilde z_{i+1}-z_j)}{(\tilde z_{j+1}-z_{i+1})(\tilde z_i-z_j)}\quad \text{for $i<j$},
\end{align*}
where $\tilde{z}_i=z_{i+n+1}=z_{i}+1$ and indices live in $\mathbb Z_{2n+2}$.

\subsection{Degenerations of $\mathscr A_n$ with perfect $u$ equations} 

The degenerations of $\mathscr A_n$ are given by setting some $C$'s in ABHY construction to zero or deleting the corresponding elements in the building set. Note that the building set after deletion doesn't need to be a new building set, \emph{i.e.} a degeneration of $\mathscr A_n$ may not be a generalized permutohedron. However, we can still ask the following question: are these degenerations binary geometries? 
Of course, we should require at least that the number of facets of the degeneration of $\mathscr A_n$ should be equal to the size of the corresponding deleted building set minus one, \emph{i.e.} its big polyhedron is a simplex.
Before answering this question, let's first try to find all degenerations of $\mathscr A_n$ with this property.


From the following equations for facets $\{F_{[i,j]}\}$ of $\mathscr A_n$
\[
	F_{[i,k]}+F_{[j,l]}-F_{[i,l]}-F_{[j,k]}=
	\sum_{i\leq a<j\atop{k<b\leq l}} C_{[a,b]} 
\]
where $F_{[a,b]}:=0$ when $a>b$ or $[a,b]=[0,n]$,
a standard way to get a degeneration whose big polyhedron is a simplex is setting all $C$'s on the RHS of 
\begin{align}
&F_{[i,j-1]}+F_{[j,l]}-F_{[i,l]}= \sum_{i\leq a<j\leq b\leq l} C_{[a,b]} 
\quad \text{for $i<j=k+1\leq l$ }\\\nonumber
&F_{[0,k]}+F_{[j,n]}-F_{[j,k]}= \sum_{0\leq a<j\atop{k<b\leq n}} C_{[a,b]} 
\quad \text{for $0<j\leq n$ and $0\leq k<n$}
\end{align}
to be zero, and all wanted degenerations can be got by setting $C$'s to zero in several equations above. 

From the viewpoint of the mesh picture, an equation above corresponds to a big diamond whose top or bottom corner and all $C$'s are zero. For example,
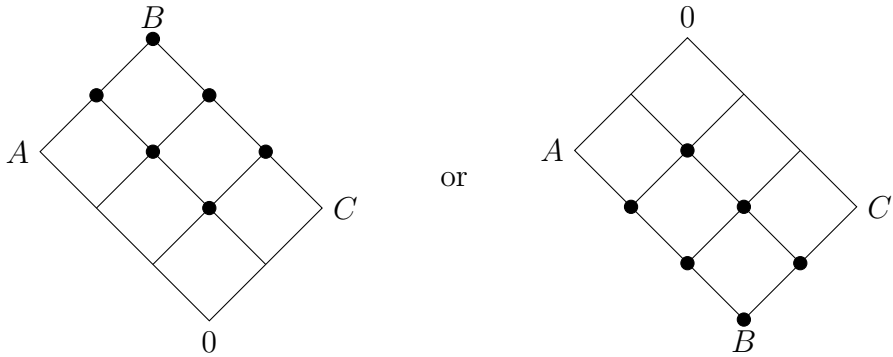
\begin{center}
\begin{tikzpicture}[scale=0.75,baseline={([yshift=-.5ex]current bounding box.center)}]
\draw (-2,0) node[left] {$A$} -- (1,-3) node[below] {$0$}
	-- (3,-1) node[right] {$C$} -- (0,2) node[above] {$B$}-- cycle;
\draw (-1,-1) -- (1,1);
\draw (-1,1) -- (2,-2);
\draw (0,-2) -- (2,0);
\node[circle,draw=black, fill=black, inner sep=0pt,minimum size=5pt] at (1,1) {};
\node[circle,draw=black, fill=black, inner sep=0pt,minimum size=5pt] at (0,2) {};
\node[circle,draw=black, fill=black, inner sep=0pt,minimum size=5pt] at (-1,1) {};
\node[circle,draw=black, fill=black, inner sep=0pt,minimum size=5pt] at (0,0) {};
\node[circle,draw=black, fill=black, inner sep=0pt,minimum size=5pt] at (1,-1) {};
\node[circle,draw=black, fill=black, inner sep=0pt,minimum size=5pt] at (2,0) {};
\end{tikzpicture}
\qquad \text{or}\qquad
\begin{tikzpicture}[scale=0.75,baseline={([yshift=-.5ex]current bounding box.center)}]
\draw (-2,0) node[left] {$A$} -- (1,-3) node[below] {$B$}
-- (3,-1) node[right] {$C$} -- (0,2) node[above] {$0$}-- cycle;
\draw (-1,-1) -- (1,1);
\draw (-1,1) -- (2,-2);
\draw (0,-2) -- (2,0);
\node[circle,draw=black, fill=black, inner sep=0pt,minimum size=5pt] at (0,0) {};
\node[circle,draw=black, fill=black, inner sep=0pt,minimum size=5pt] at (1,-3) {};
\node[circle,draw=black, fill=black, inner sep=0pt,minimum size=5pt] at (-1,-1) {};
\node[circle,draw=black, fill=black, inner sep=0pt,minimum size=5pt] at (0,-2) {};
\node[circle,draw=black, fill=black, inner sep=0pt,minimum size=5pt] at (2,-2) {};
\node[circle,draw=black, fill=black, inner sep=0pt,minimum size=5pt] at (1,-1) {};
\end{tikzpicture}
\end{center}
we can see that
\[
A+C-B=0.
\]
Since $A,B,C\geq 0$, the above equation forces $A$ and $C$ to vanish when $B=0$. Therefore, $B$ is no longer a facet of the degeneration. Similarly, black points are all vanishing facets in the above example, and generally
the number of vanishing facets equals the number of deleted $C$'s for degenerations of this type.

One can use these diamonds to build a degeneration with more vanishing facets whose big polyhedron is a simplex. For example, in $\mathscr A_3$
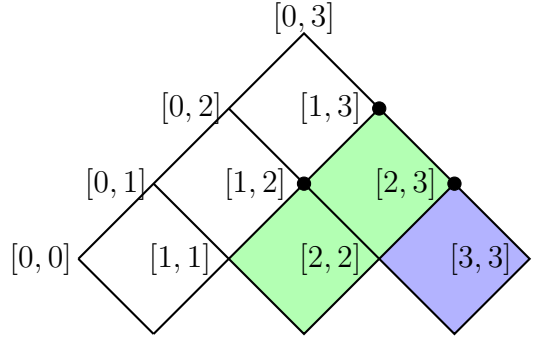
\begin{center}
\begin{tikzpicture}
\fill[green!30] (2,1) -- (0,-1) -- (1,-2) -- (3,0);
\fill[blue!30] (3,0) -- (2,-1) -- (3,-2) -- (4,-1);
\draw[thick] (-2,-1) -- (1,2) -- (4,-1) -- (3,-2) -- (0,1);
\draw[thick] (-1,-2) -- (2,1);
\draw[thick] (-2,-1) -- (-1,-2);
\draw[thick] (-1,0) -- (1,-2) -- (3,0);
\node[circle,draw=black, fill=black, inner sep=0pt,minimum size=5pt] at (1,0) {};
\node[circle,draw=black, fill=black, inner sep=0pt,minimum size=5pt] at (2,1) {};
\node[circle,draw=black, fill=black, inner sep=0pt,minimum size=5pt] at (3,0) {};
\node at (-2.5,-1) {$[0,0]$};
\node at (-0.65,-1) {$[1,1]$};
\node at (1.35,-1) {$[2,2]$};
\node at (3.35,-1) {$[3,3]$};
\node at (-1.5,0) {$[0,1]$};
\node at (0.35,0) {$[1,2]$};
\node at (2.35,0) {$[2,3]$};
\node at (-0.5,1) {$[0,2]$};
\node at (1.35,1) {$[1,3]$};
\node at (1,2.2) {$[0,3]$};
\end{tikzpicture}
\end{center}
after setting $C$'s in green and purple diamonds to be zero, three facets (black points) vanish and building set after deletion is 
\[
	\{[0,3],[0,2],[0,1],[0,0],[1,1],[2,2],[3,3]\}.
\]
One can further check that it's a cube (or $\mathscr A_1^3$).

We can play the game for any degenerations of $\mathscr A_n$, count black points and deleted $C$'s in each degeneration, and find that for $\mathscr A_2$, $\mathscr A_3$ and $\mathscr A_4$, there're respectively $4$, $41$ and $580$ possible degenerations whose big polyhedra are simplices. For these degenerations, we can easily find their $u$ variables if the set of all vanishing facets are known.

Suppose $\mathscr D$ is the set of all vanishing facets of a degeneration in the building set, 
we can read from the mesh that
\[
F_{I'}=\sum_{I\not\in \mathscr D}M_{I'}^IF_I
\]
for any $I'\in \mathscr D$, where $M$ is a matrix with non-negative entries. Therefore,
\[
\prod_{I\in \mathscr A_n}u_{I}^{F_{I}}=
\prod_{I'\in \mathscr D}u_{I'}^{F_{I'}}
\cdot
\prod_{I\not\in \mathscr D}u_{I}^{F_{I}}=
\prod_{I\not\in \mathscr D}\biggl(
u_I\prod_{I'\in \mathscr D}u_{I'}^{M_{I'}^I}
\bigg)^{F_I},
\]
so the new $u$ variables of the degeneration are
\begin{equation}
    \tilde u_{I}=u_I\prod_{I'\in \mathscr D}u_{I'}^{M_{I'}^I}
\end{equation}
for any $I\not\in \mathscr D$. 

The simplest degenerations of $\mathscr A_n$ are those with one $C_{[i,i+1]}$ vanishing, where we get 
\[
	F_{[i,i+1]}=F_{i}+F_{i+1},
\]
with $F_i=F_{[i,i]}$, and we find the $u$ variables become
\[
	\tilde u_{i}=u_{i} u_{[i,i+1]},\quad 
	\tilde u_{i+1}=u_{i+1} u_{[i,i+1]},\quad 
	\tilde u_I=u_I \text{ for the other $I\neq [i,i+1]$}.
\]
They satisfy the following $u$ equation
\begin{equation}
	1-\tilde u_{i}=\prod_{0\leq a\leq i-1}\tilde u_{[a,i-1]}
	\prod_{i+1< b\leq n} \tilde u_{[i+1,b]},\quad 
	1-\tilde u_{i+1}=\prod_{0\leq a <i-1} \tilde u_{[a,i]}
	\prod_{i+2\leq b\leq n} \tilde u_{[i+2,b]},\quad 
\end{equation}
and the other $u$ equations are original $u$ equations for $\{u_{[i,j]}\,:\, [i,j]\neq [i,i],[i,i+1],[i+1,i+1]\}$ in terms of new variables. Similarly, one can easily find the $u$ variables and $u$ equations for the degenerations with $C_{[i,i+1]}=0$ for several $i$.


The next simplest degenerations of $\mathscr A_n$ may be products 
$\mathscr A_i\times \mathscr A_{n-i}$, which can be realized as the following
mesh,
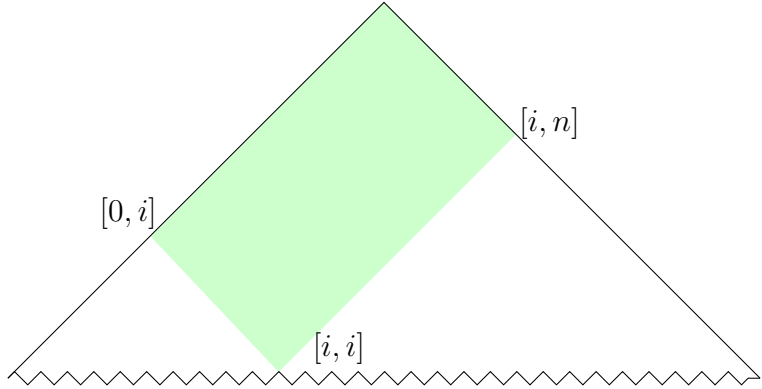
\begin{center}
\usetikzlibrary{decorations.pathmorphing}
\begin{tikzpicture}[scale=2]
\fill[green!20] (0.4498,-0.558) -- (1.2997,-1.4584) -- (2.8779,0.1223) --(2,1);
\draw (-0.5,-1.5) -- (2,1) -- (4.5,-1.5);
\draw[decorate,decoration=zigzag] (-0.5,-1.5)--(4.5,-1.5);
\node at (1.7,-1.3) {$[i,i]$};
\node at (0.3,-0.4) {$[0,i]$};
\node at (3.1,0.2) {$[i,n]$};
\end{tikzpicture}
\end{center}
where $C$'s in green region vanish, \emph{i.e.} $C_{[a,b]}=0$ for $0\leq a<i<b\leq n$.
Vanishing facets of this degeneration are
\[
F_{[a,b]}=F_{[a,n]}+F_{[0,b]}\quad \text{for $0<a\leq i\leq b<n$},
\]
so new $u$ variables are
\begin{equation}
	\tilde u_{[0,b]} = u_{[0,b]}\prod_{a=1}^{i} u_{[a,b]},\quad 
	\tilde u_{[a,n]} = u_{[a,n]}\prod_{b=i}^{n-1} u_{[a,b]},
\end{equation}
and $\tilde u_{I}=u_{I}$ for the other $I$. The $u$ equations are the union of 
$u$ equations of the following two independent meshes
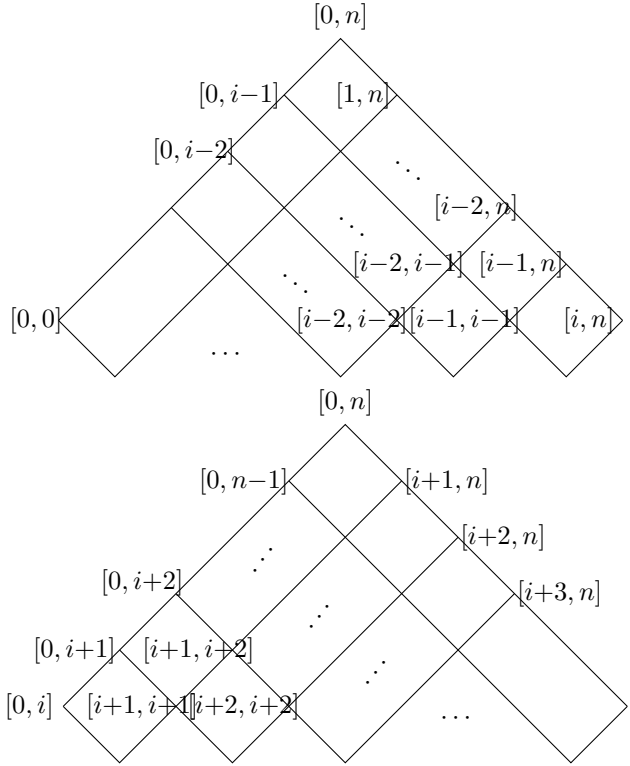
\begin{center}\footnotesize
\begin{tikzpicture}[scale=1.5,baseline={([yshift=-.5ex]current bounding box.center)}]
\draw (0.5,-2.5) -- (3,0) -- (5.5,-2.5);
\draw (5.5,-2.5) -- (5,-3) -- (2.5,-0.5);
\draw (5,-2) -- (4,-3) -- (2,-1);
\draw (4.5,-1.5) -- (3,-3) -- (1.5,-1.5);
\draw (0.5,-2.5) -- (1,-3) -- (3.5,-0.5);
\node at (2,-2.8) {$\cdots$};
\node at (2.6,-2.1) {$\ddots$};
\node at (3.1,-1.6) {$\ddots$};
\node at (3.6,-1.1) {$\ddots$};
\node at (0.3,-2.5) {$[0,0]$};
\node at (3.1,-2.5) {$[i{-}2,i{-}2]$};
\node at (4.1,-2.5) {$[i{-}1,i{-}1]$};
\node at (5.2,-2.5) {$[i,n]$};
\node at (4.6,-2) {$[i{-}1,n]$};
\node at (3.6,-2) {$[i {-}2,i{-}1]$};
\node at (4.2,-1.5) {$[i{-}2,n]$};
\node at (3.2,-0.5) {$[1,n]$};
\node at (3,0.2) {$[0,n]$};
\node at (2.1,-0.5) {$[0,i{-}1]$};
\node at (1.7,-1) {$[0,i{-}2]$};
\end{tikzpicture}\quad 
\begin{tikzpicture}[scale=1.5,baseline={([yshift=-.5ex]current bounding box.center)}]
\draw (0.5,-2.5) -- (3,0) -- (5.5,-2.5);
\draw (5.5,-2.5) -- (5,-3) -- (2.5,-0.5);
\draw (4.5,-1.5) -- (3,-3) -- (1.5,-1.5);
\draw (4,-1) -- (2,-3) -- (1,-2);
\draw (0.5,-2.5) -- (1,-3) -- (3.5,-0.5);
\node at (4,-2.6) {$\cdots$};
\node[rotate=90] at (3.2,-2.2) {$\ddots$};
\node[rotate=90] at (2.7,-1.7) {$\ddots$};
\node[rotate=90] at (2.2,-1.2) {$\ddots$};
\node at (3,0.2) {$[0,n]$};

\node at (3.9,-0.5) {$[i{+}1,n]$};
\node at (4.4,-1) {$[i{+}2,n]$};
\node at (4.9,-1.5) {$[i{+}3,n]$};
\node at (2.1,-0.5) {$[0,n{-}1]$};
\node at (1.2,-1.4) {$[0,i{+}2]$};
\node at (0.2,-2.5) {$[0,i]$};
\node at (0.6,-2) {$[0,i{+}1]$};
\node at (1.2,-2.5) {$[i{+}1,i{+}1]$};
\node at (1.7,-2) {$[i{+}1,i{+}2]$};
\node at (2.1,-2.5) {$[i{+}2,i{+}2]$};
\end{tikzpicture}
\end{center}
One could use this construction of products to produce any finite product $\mathscr A_{i_1}\times \cdots \times \mathscr A_{i_k}$.
The direct proof of this argument for $\mathscr A_n$ is using extended $u$ equations \cite{Brown:2009qja,Arkani-Hamed:2019plo} of $\mathscr A_n$. 

We can generalize the simplest cases in another way:
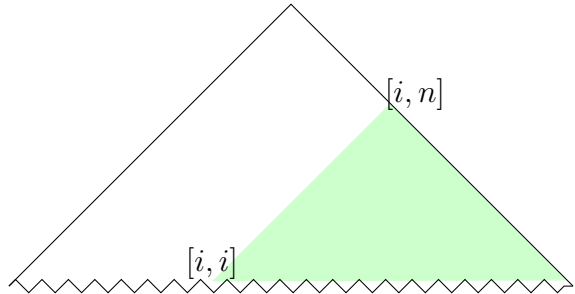
\begin{center}
\begin{tikzpicture}[scale=1.5]
\fill[green!20] (4.44,-1.45) -- (1.3,-1.4584) -- (2.88,0.12) --(2,1);
\draw (-0.5,-1.5) -- (2,1) -- (4.5,-1.5);
\draw[decorate,decoration=zigzag] (-0.5,-1.5)--(4.5,-1.5);
\node at (1.3,-1.3) {$[i,i]$};
\node at (3.1,0.2) {$[i,n]$};
\end{tikzpicture}
\end{center}
setting 
\[
C_{[k,l]}=0 \quad \text{for $0<i\leq k< l\leq n$}.
\]
In this case, 
\[
F_{[k,l]}=\sum_{m=k}^l F_{[m,m]}
\]
for $i\leq k < l \leq n$. Therefore,
\[
\tilde u_{[k,k]}=\prod_{i\leq p\leq k\leq q \leq n}u_{[p,q]}=\frac{x_k}{x_{[i-1,k]}}\quad 
\text{for $k\in [i,n]$},\quad \tilde u_I=u_I\quad \text{for the other $I$},
\]
and their $u$ equations are
\begin{equation}
1-\tilde u_{[k,k]}=\prod_{j=0}^{i-1}\tilde u_{[j,k-1]},
\end{equation}
and the other $u$ equations are old $u$ equations in terms of new variables.
When $i=1$, the degeneration is $\mathscr A_1^n$.

Similarly, we can also consider the following degenerations:
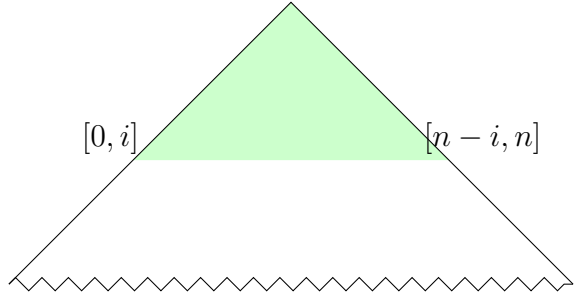
\begin{center}
\begin{tikzpicture}[scale=1.5]
\fill[green!20] (3.4,-0.4) -- (2,-0.4) -- (0.6,-0.4) --(2,1);
\draw (-0.5,-1.5) -- (2,1) -- (4.5,-1.5);
\draw[decorate,decoration=zigzag] (-0.5,-1.5)--(4.5,-1.5);
\node at (0.4,-0.2) {$[0,i]$};
\node at (3.7,-0.2) {$[n-i,n]$};
\end{tikzpicture}
\end{center}
setting
\[
C_{[k,l]}=0 \quad \text{for $l-k>i$}.
\]
In this case,
\[
F_{[k,l]}=F_{[0,l]}+F_{[k,n]}\quad \text{for $l-k\geq i$, $1\leq k\leq n-i-1$ and $i+1\leq l\leq n-1$},
\]
then 
\[
\tilde u_{[0,l]}=\prod_{j=0}^{l-i+1}  u_{[j,l]},\quad 
\tilde u_{[k,n]}=\prod_{j=k+i-1}^{n}  u_{[k,j]}.
\]
Their $u$ equations are 
\begin{equation}
1-\tilde u_{[0,l]}=\prod_{a=l-i+2}^{l+1}\tilde u_{[a,n]}\cdot\prod_{a=l-i+2}^{l+1}\prod_{b=l+1}^{a+i-2} \tilde u_{[a,b]},\quad
1-\tilde u_{[k,n]}=\prod_{b=k-1}^{k+i-2}\tilde u_{[0,b]}\cdot \prod_{b=k-1}^{k+i-2}\prod_{a=\max(b-i+2,1)}^{k-1} \tilde u_{[a,b]}.
\end{equation}
When $i=1$, we get $\mathscr A_1^n$ again.

\subsection{Degenerations of $\mathscr B_n$ and products}
\subsubsection{Degenerations of $\mathscr B_n$}

The general treatment of degenerations of $\mathscr B_n$ is more difficult than $\mathscr A_n$ because it has a long ABHY condition eq.\eqref{abhyB}. Here we only show some examples of degenerations of $\mathscr B_n$ with perfect $u$ equations.

The first degeneration of $\mathscr B_n$ with perfect $u$ equations is $\mathscr A_n$ because the building set of $\mathscr A_n$ is contained in the building set of $\mathscr B_n$.

The simplest degenerations of $\mathscr B_n$ are those with $C_{[i,i+1]}=0$. For which we get,
\[
	F_{[i,i+1]}=F_{i}+F_{i+1},
\]
where $F_i=F_{[i,i]}$, so 
\[
	\tilde u_{i}=u_{i} u_{[i,i+1]},\quad 
	\tilde u_{i+1}=u_{i+1} u_{[i,i+1]},\quad 
	\tilde u_I=u_I \text{ for the other $I\neq [i,i+1]$}.
\]
The $u$ equations of this degeneration are
\[
	1-\tilde u_i=\tilde u_{[i+1,i-1]}^2\tilde u_{[i-1,i-1]}
	\prod_{a=i+2}^{i-2}\tilde u_{[a,i-1]}\tilde u_{[i+1,a]},\quad
	1-\tilde u_{i+1}=
	\tilde u_{[i+2,i]}^2\tilde u_{[i+2,i+2]}\prod_{a=i+3}^{i-1}\tilde u_{[a,i]}\tilde u_{[i+2,a]},
\]
and the other $u$ equations are old $u$ equations in terms of new $u$ variables.

The above degeneration corresponds to deleting an element in $\mathscr B_n$, and the following degeneration corresponds to adding an element to $\mathscr A_n$. 

Consider the building set obtained by adding the set $K_i=[0,i]\cup [i+2,n]$ to $\mathscr A_n$. It is also a degeneration of $\mathscr B_n$ by setting the other $C$'s to be zero. It's easy to solve equations of the facets and get that 
\[
C_{K_i}=F_{[0,i]}+F_{[i+2,n]}-F_{K_i},\quad 
C_{[0,n]}=F_{K_i}+F_{[0,n-1]}+F_{[1,n-1]}-F_{[1,n-1]}-F_{[0,i]}-F_{[i+2,n]},
\]
the other $C_I$ are given by eq.\eqref{CinFA}. Therefore, new $u$ variables are
\[
\tilde u_{K_i}=\frac{x_{K_i}}{x_{[0,n]}},\quad 
\tilde u_{[0,i]}=\frac{x_{[0,n]}}{x_{K_i}}u_{[0,i]},\quad 
\tilde u_{[i+2,n]}=\frac{x_{[0,n]}}{x_{K_i}}u_{[i+2,n]},
\]
and the other $\tilde u_I$ is equal to $u_I$. The $u$ equations are
\[
1-\tilde u_{K_i}=\prod_{I\ni i+1}\tilde u_I,\quad
1-\tilde u_{[0,i]}=\prod_{J}\tilde u_J^{J||[0,i]}\prod_{K\supset [i+2,n]}\tilde u_K,\quad 
1-\tilde u_{[i+2,n]}=\prod_{J}\tilde u_J^{J||[i+2,n]}\prod_{K\supset [0,i]}\tilde u_K, 
\]
where $I||J$ are compatability degree of $I$ and $J$ in $\mathscr A_n$,
and the other $u$ equations are corresponding $u$ equations for $\mathscr A_n$ in new variables. 

For example, adding $K_0$ to $\mathscr A_2$ we get $\mathscr B_2$, and  adding $K_0$ to $\mathscr A_3$ gives the following $u$ equations
\begin{align*}
1-\tilde u_{3}&=\tilde u_{2} \tilde u_{12} \tilde u_{012}&
1-\tilde u_{01}&=\tilde u_{2} \tilde u_{12} \tilde u_{23} \tilde u_{023} \tilde u_{123}\\
1-\tilde u_{1}&=\tilde u_{0} \tilde u_{2} \tilde u_{23} \tilde u_{023}^2&
1-\tilde u_{2}&=\tilde u_{1} \tilde u_{3} \tilde u_{01}\\
1-\tilde u_{012}&=\tilde u_{3} \tilde u_{23} \tilde u_{023} \tilde u_{123}&
1-\tilde u_{023}&=\tilde u_{1} \tilde u_{01} \tilde u_{12} \tilde u_{012} \tilde u_{123}\\
1-\tilde u_{123}&=\tilde u_{0} \tilde u_{01} \tilde u_{012} \tilde u_{023}&
1-\tilde u_{0}&=\tilde u_{1} \tilde u_{12} \tilde u_{23} \tilde u_{123}^2\\
1-\tilde u_{12}&=\tilde u_{0} \tilde u_{3} \tilde u_{01} \tilde u_{23} \tilde u_{023}^2&
1-\tilde u_{23}&=\tilde u_{0} \tilde u_{1} \tilde u_{01}^2 \tilde u_{12} \tilde u_{012}^2.
\end{align*}

More generally we can consider adding a triangle namely $\cup_{l=0}^{k-1}\cup_{m=l}^{k-1} K_{i,l,m}$ with $K_{i,l,m}=\{[0,i+l]\cup [i+2+m,n] \}$ and $k=1,\dots,(n-1)$ to the building set of $\mathscr{A}_n$.
For $k=1$ we get the previous example and for $k=n-1$ we get ${\mathscr B_n}$.

Without loss of generality we shall consider only the $i=0$ case and we shall denote $K_{0,l,m}$ by just $K_{l,m}$.
The $u$ variables are
\bea
\tilde u_{K_{l,m}}=\begin{cases} \frac{x_{K_{l,m}}}{x_{[0,n]}},~~~~~~ {\rm if} ~m=l \\
\frac{x_{K_{l,m}} x_{K_{l+1,m-1}}}{x_{K_{l+1,m}}x_{K_{l,m-1}}}
,~ m\neq l \end{cases}\quad 
\tilde u_{[0,l]}=\frac{u_{[0,l]}}{\prod_{j=0}^{k-l-1} \tilde u_{K_{l,l+j}}},\quad 
\tilde u_{[m+2,n]}=\frac{u_{[m+2,n]}}{\prod_{j=0}^{m}\tilde u_{K_{j,m}}},
\eea
with the  understanding that $K_{1,0}=K_{2,0}=[0,n]$ and the other $\tilde u_I$ is equal to $u_I$. 

The $u$ equations are:
\begin{align*}
&1-\tilde u_{[0,l]}=\left(\prod_{i=0}^{l-1} \prod_{j=i+1}^{k+1}  \tilde u_{[0,i]\cup [j,n]} \prod_{j=i+1}^{l+1} \tilde u_{[0,i]\cup [j,n]}\right) \left(\prod_{i=1}^{l+1} \prod_{j=l+1}^{n}\tilde u_{[i,j]} \right) \left(\prod_{i=1}^{k+1} \tilde u_{[i,n]} \right),\\[2ex]
&1-\tilde u_{[m+2,n]}=\left( \prod_{j=0}^{m}\tilde u_{[0,j]} \prod_{i=1}^{n-m-1}\tilde u_{[0,m+i]}^2 \right) \left( \prod_{i=m+1}^{n-1}\prod_{j=1}^{m+1}\tilde u_{[j,i]} \right) \left(\prod_{j=0}^{k-m-2} \prod_{i=1}^{m} \tilde u_{[0,i]\cup [3+j+m, n]} \prod_{i= m+1}^{j+m+1} \tilde u_{[0,i]\cup [3+j+m,n]}^2 \right) \\[2ex]
&1-\tilde u_{K_{l,m}}=\left(\prod_{a=0}^1\prod_{i=l+1}^{n}\tilde u_{[a,i]}\right) \left(\prod_{{\tiny {K_{l',m'} \not\subset K_{l,m}}\atop{K_{l,m} \not\subset K_{l',m'}} }} \tilde u_{K_{l',m'}}^{(l',m')||(l,m)}\right)\left(\prod_{j=2}^{m+1} \prod_{i=l+1}^{n} \tilde u_{[j,i]} \right)\times 
\begin{cases}
 1, ~{\rm if} ~ l=m  \\
 \prod_{i=0}^{l+1} \prod_{j=m+1}^n \tilde u_{[i,j]},~ {\rm if}~l \neq m
\end{cases}
\end{align*}
where $(l',m')||(l,m)$ is the compatability degree of $K_{l,m}$ with respect to $K_{l',m'}$ in $\mathscr B_n$ with the understanding that $\tilde u_{[0,n]}=1$ and the other $u$ equations are corresponding $u$ equations for $\mathscr A_n$ in new variables.



\subsubsection{Products of ${\mathscr A}$ and $\mathscr{B}$}
We can also construct binary geometries of the type $\prod_{i=1}^n X_i$ where $X_i={\mathscr A_i}~{\rm or}~{\mathscr B_i}$ which are not necessarily degenerations of ${\mathscr A}~{\rm or}~{\mathscr B}$. Without loss of generality we consider only the case ${\mathscr A_n}\times {\mathscr B_m}$. We can do this in a couple of simple ways as follows:

(1) We consider the building set $\mathbf{B}_{1}\cup \mathbf{B}_{2}$ with $\mathbf B_1= \{[i,j]\,:\, 0\leq i\leq j\leq n\}$ and $\mathbf B_2= \{[i,i+k]\,:\, n\leq i, k \leq m+n\}$. Notice that these are just the building sets of ${\mathscr A_n}$ and ${\mathscr B_m}$ respectively.

If $\mathbf{B}_{1}$ and $\mathbf{B}_{2}$ were disjoint then it's obvious that the stringy integral ${\cal I}_{\mathbf B}$ in \eqref{stringycanonicalform} directly factorizes into ${\cal I}_{\mathbf B_1}\times {\cal I}_{\mathbf B_2}$, however the singlet $\{n\}$ belongs to both $B_1$ and $B_2$. But by rescaling the variables in $B_2$ by $x_i \rightarrow x_n x_i$ for all $i=n+1,\dots,n+m $ it is straightforward to see that the ${\cal I}_{\mathbf B}$ does indeed factorize into ${\cal I}_{\mathbf B_1}\times {\cal I}_{\mathbf B_2}$ with $x_n$ playing the role of $x_0=1$ in $B_2$.

Notice that the above proof did not use the fact that ${\mathbf B_1}, {\mathbf B_2}$ were building sets of ${\mathscr A},~{\mathscr B}$ and can easily be carried forward to obtain any product of generalized permutohedra. The interested reader may refer appendix (\ref{appendix:A}) for details.

(2) Consider the building set  $\mathbf{B}_{1}\cup \mathbf{B}_{2}$ with $\mathbf{B}_{1}=\{[i,i{+}k]\,:\,0\leq i, k\leq m\}$ and $\mathbf{B}_{2}=\{[i,j]: m<i<j\leq n+m \}\cup \{[0,i]:  m\leq i\leq n+m\}$. $\mathbf{B}_{1}$ is obviously the building set for $\mathscr{B}_{m}$, while $\mathbf{B}_{2}$ can be viewed as the building set for $\mathscr{A}_{n}$ by identifying $[0,n]$ as a \emph{singleton} in $\mathbf{B}_{2}$. Due to this identification, one can easily write down the $u$ variables separately for $\mathbf{B}_{1}$ and $\mathbf{B}_{2}$, and check that these $u$ variables satisfy the $u$ equations for $\mathscr{A}_{n}\times \mathscr{B}_{m}$.

\section{Stringy canonical forms and binary geometries for generalized permutohedra}
In this section we shall argue that a large class of generalized permutohedra which are realised as degenerations of permutohedron ${\mathscr P_n}$ are binary geometries. Before we proceed, we shall review some details about the generalized permutohedra \cite{Postnikov:2005,Postnikov:2006} which we shall use throughout the paper.
\subsection{Generalized permutohedra}\label{GP}
A generalized permutohedron is a polytope that can be obtained as Minkowski sums and differences of coordinate simplices. 

Let $\Delta_{[0,n]} = \operatorname{ConvexHull}(e_1,\dots,e_n)$ be the standard coordinate simplex in $\mathbb{R}^{n+1}$. Then we have:
\bea
{\mathscr P_n}(\{y_I \})= \sum y_I . \Delta_I  
\eea
for some collection of subsets $I \in [0,n]$.
We shall restrict ourselves to the large class of generalized permutohedra which admit such a realisation only as Minkowski sums except translations i.e, $y_I \ge 0$ for all non-singlets $I$~\footnote{This is mainly due to the subtle nature of the Minkowski difference operation, which makes the shape of the resulting polytope dependent on the relative magnitudes of $y_I$'s. We allow $y_I$ for singlets to be  negative as it is only a translation and to connect with \eqref{intgenperm}.} which are called nestohedra. Throughout the paper we shall mean nestohedra or products of nestohedra whenever we refer to generalized permutohedra.

The combinatorial structure of the generalized permutohedron ${\mathscr P_n(\{y_I\})}$ (nestohedra) is independent of $y_I$ and depends only on the collection of subsets $I \in [0,n]$ of the  building set ${\mathbf B}$, which is 
a collection of non-empty subsets of $[0,n]$ satisfying the following: 
\begin{compactenum}[\quad (1)]
    \item If $I,J \in {\mathbf B}$ and $I \cap J \neq \varnothing $, then $I \cup J \in {\mathbf B}$.
    \item ${\mathbf B}$ contains all singletons $\{ i\}$ for $i \in S$.
\end{compactenum}
Here are some interesting examples of generalized permutohedra:
\begin{itemize}
    \item  If ${\mathbf B}= \{[i] | i=0,\dots,n \}$ is the complete flag of intervals, then ${\mathscr P_n(\{y_I\})}$ is the Stanley-Pitman polytope or Hypercube.
    \item If ${\mathbf B}$ corresponds to all the non empty subsets of $[0,n]$ and $Y_I =y_{|I|}$ {\it i.e.}, the variables $Y_I$ are equal for all subsets of the same cardinality, then ${\mathscr P_n(\{y_I\})}$ is the usual permutohedron ${\mathscr P_n}$. 
    \item If ${\mathbf B}=\{ [i,j] | 0\leq  i \leq j \leq n\}$ is the set of consecutive intervals, then ${\mathscr P_n(\{y_I\})}$ is the associahedron ${\mathscr A_n}$. 
    \item If  $\mathbf B=\{[i,i+k]\,:\,0\leq i, k\leq n\}$ is the set of cyclic intervals, then ${\mathscr P_n(\{y_I\})}$ is the cyclohedron ${\mathscr B_n}$.
    \item Let $\Gamma$ be a graph on the vertex set $[0,n]$. Let us assume that ${\mathbf B} = {\mathbf B}(\Gamma)$ is the set of subsets $I \in [0,n]$ such that the induced graph $\Gamma$ is connected, then  ${\mathscr P_n(\{y_I\})}$ is a graph associahedron.
\end{itemize}

\paragraph{Nested complex} To describe the combinatorial structure we need the notion of {\it nested complex}. We shall only summarize the necessary results here and let the reader interested in the details to refer \cite{Postnikov:2005,Postnikov:2006}.
A subset ${\mathbf N}$ in the building set ${\mathbf B}$ is called a {\it nested set} if it satisfies the following conditions:
\begin{compactenum}[\quad (1)]
    \item For any $I,J \in {\mathbf N}$, we either have $I \subset J$ or $J\subset I$ or $I$ and $J$ are disjoint.
    \item For any collection of $k \geq 2$ disjoint subsets $J_1,J_2,\dots, J_k \in \mathbf N$ their union $J_1 \cup \cdots \cup J_k$ is not in $\mathbf B$.
    \item ${\mathbf N}$ contains all maximal elements of $\mathbf B$.
\end{compactenum}
The {\it nested complex} ${\mathbf N}({\mathbf B})$ is defined as the poset of the set of all nested sets in ${\mathbf B}$ ordered by inclusion.
\paragraph*{Facial structure} 
Let us assume that the set ${\mathbf B}$ associated with a generalized permutohedron ${\mathscr P_n(\{y_I\})}$ is a building set on $[0,n]$. \emph{Then the poset of faces ${\mathscr P_n(\{y_I\})}$ ordered by reverse inclusion is isomorphic to the nested complex ${\mathbf N}({\mathbf B})$}. 

In particular, it means that (1) the set of facets of the generalized permutohedron just corresponds to the set of all elements in ${\mathbf B}$ excluding the maximal element, and (2) two facets corresponding to the sets $I$ and $J$ in ${\mathbf B}$ intersect {\it i.e.}, are {\it compatible} with each other, if and only if  either $I \subset J$,~ $J \subset I$~ or $I \cap J = \varnothing ~{\text and} ~ I \cup J \notin {\mathbf B}$.

The face $P_\mathbf{N}$ of ${\mathscr P_n(\{y_I\})}$ associated with the nested set $\mathbf{N} \in \mathbf{N}(\mathbf{B})$ is given by
\bea\label{deffacet}
P_{\mathbf{N}} = \bigl\{ (t_0,\dots,t_{n}) \in \mathbb{R}^{n+1} | \sum_{i \in I} t_i =z_{I}~{\rm for}~I \in \mathbf{N}; ~ \sum_{i \in J}~t_i \geq z_J,~ {\rm for} ~J \in \mathbf{B}  \bigr\} \:,
\eea
where $z_I=\sum_{J\subset I,J\in\mathbf B}y_J$ for $I\in \mathbf B$ and coordinate simplices $\{\Delta_I\}_{I\in\mathbf B}$ are living in $\{t_i\}_{0\leq i\leq n}$-space.
In particular, the equation of the facet labeled by nested set $\{[0,n],I\}$ of ${\mathscr P_n(\{y_I\})}$ is given by
\begin{equation}\label{eqoffacet}
\sum_{i\in I}t_i=\sum_{J\subset I,J\in\mathbf B}y_J.
\end{equation}

Another description of the faces is the following: for each decomposition $[0,n]=  \sqcup_{I \subset {\mathbf N} } S_I $ where $S_I$ are non-empty, the face $P_\mathbf{N}$ of ${\mathscr P_n(\{y_I\})}$ associated with the nested set ${\mathbf N} \in {\mathbf N}({\mathbf B})$ is
\bea
P_\mathbf{N} = \sum_{{J \subset {\mathbf B}}\atop{J \cap S_I \neq \varnothing}} y_{J} \Delta_{J \cap S_I} \:.
\eea
In particular each facet of $\mathscr{P}_{n}(\{y\})$ corresponding to a set $I \in \mathbf{B}$ is isomorphic to the product $\mathscr{P}_{ \mathbf{B}_{I}} \times \mathscr{P}_{\mathbf{B}_{\bar{I}}}$ of generalized permutohedra corresponding to the building sets 
\begin{equation}\label{BI}
    \mathbf{B}_{I} = \{J \vert J\subset I, J\in \mathbf{B}\}
\end{equation}
and
\begin{equation} \label{BIbar}
    \mathbf{B}_{\bar{I}}=\{J-I \vert J\in \mathbf{B}\}
\end{equation}




Let us look at a couple of examples to appreciate these facts better: \\
(1) Consider the 2d associahedron with building set 
\[{\mathbf B}= \{ \{0\},\{1\},\{2\},\{0,1\},\{1,2\},\{0,1,2\} \}.\]
This polygon has 5 facets which correspond to the sets $\{ \{0\},\{1\},\{2\},\{0,1\},\{1,2\}\}$ and their compatability is shown in the figure below. \\
(2) Consider the 2d cyclohedron/permutohedron with building set 
\[{\mathbf B}= \{ \{0\},\{1\},\{2\},\{0,1\},\{1,2\},\{0,2\},\{0,1,2\} \}.\]
This polygon has 6 facets which correspond to the sets $\{ \{0\},\{1\},\{2\},\{0,1\},\{1,2\},\{0,2\}\}$ and their compatability is shown in the figure below.


\begin{figure}[H]
\begin{center}
\begin{tikzpicture}[x=0.75pt,y=0.75pt,yscale=-1,xscale=1]

\draw   (332.23,134.19) -- (310.06,198.57) -- (241.97,197.38) -- (222.06,132.26) -- (277.84,93.2) -- cycle ;
\draw   (533.33,149) -- (502.83,201.83) -- (441.83,201.83) -- (411.33,149) -- (441.83,96.17) -- (502.83,96.17) -- cycle ;

\draw (401.67,176.33) node [anchor=north west][inner sep=0.75pt]   [align=left] {01};
\draw (461,212) node [anchor=north west][inner sep=0.75pt]   [align=left] {0};
\draw (412,109) node [anchor=north west][inner sep=0.75pt]   [align=left] {1};
\draw (463,78) node [anchor=north west][inner sep=0.75pt]   [align=left] {12};
\draw (527,101) node [anchor=north west][inner sep=0.75pt]   [align=left] {2};
\draw (523,179) node [anchor=north west][inner sep=0.75pt]   [align=left] {02};
\draw (268,204) node [anchor=north west][inner sep=0.75pt]   [align=left] {0};
\draw (226,96) node [anchor=north west][inner sep=0.75pt]   [align=left] {1};
\draw (310,100) node [anchor=north west][inner sep=0.75pt]   [align=left] {12};
\draw (336,163) node [anchor=north west][inner sep=0.75pt]   [align=left] {2};
\draw (203,160) node [anchor=north west][inner sep=0.75pt]   [align=left] {01};
\draw (262,235) node [anchor=north west][inner sep=0.75pt]   [align=left] {(a)};
\draw (454,235) node [anchor=north west][inner sep=0.75pt]   [align=left] {(b)};
\end{tikzpicture}
\end{center}
\caption{Facets of the 2d associahedron (a) and permutohedron (b) as generalized permutohedra}
\end{figure}
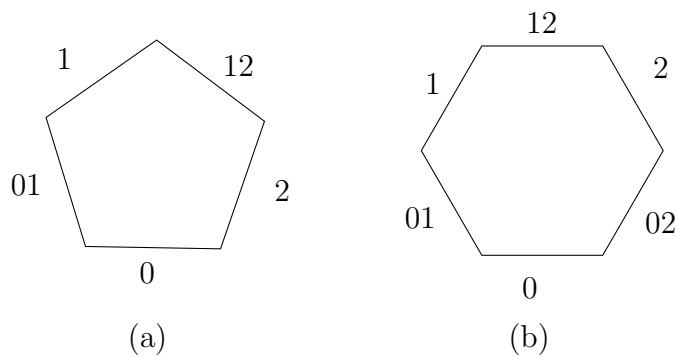

\vspace{1ex}

The big polyhedron of generalized permutohedron is a simplex. Thus it's quite natural to wonder if the associated $u$ variables satisfy some kind of $u$ equations. We claim and we shall show in the next section that this is indeed the case. Further we show that all these polytopes are binary geometries even though some of them do not admit perfect $u$ equations.

\subsection{ABHY-like realizations and $u$ variables for generalized permutohedra}

As mentioned earlier, one can write down a natural stringy integral \eqref{intgenperm} for generalized permutahedron $\mathscr{P}(\bf B)$,
and its polytope is given by the Minkowski sum
\[
	\mathscr P_n(\{-S_I\,|\,I\in \mathbf B,\,|I|>1\}) =- \sum_{I\in\mathbf{B},|I|>1}S_I.\Delta_I,
\]
where $\Delta_I=\operatorname{ConvexHull}
(\{e_i\}_{i\in I})$ is the coordinate simplex in $(S_0,\dots,S_n)$-space, so the equation of the facet labeled by $I$ is given by eq.\eqref{eqoffacet}:
\begin{equation}\label{GPfacet}
	F_I:=\sum_{J\subset I,J\in\mathbf{B}}S_J=0.
\end{equation}
From the previous subsection we know that the big polyhedron of a generalized permutohedron is a simplex.
Therefore, we can write down the matrix $W^J_A$ of the facets and obtain  
\[
S_J=F_A(W^{-1})^A_J
\]
and 
its $u$ variables 
\[ u_A=\prod_J x_J^{(W^{-1})^A_J}. \] 

Within generalized permutohedra, the original permutohedron is quite simple and important because any other generalized permutohedron can be obtained as a degeneration of the permutohedron.
The building set of the $n$-dimensional permutohedron $\mathscr P_n$ is 
the set of all non-empty subsets of $[0,n]$. 
To get $u$ equations, we should solve for $S_J$ from $F_J$ in eq.\eqref{GPfacet}, the 
solution for the permutohedron $\mathscr P_n$ is 
\begin{equation}\label{SinF}
	S_I=\sum_{J\subset I} (-1)^{|J|-|I|}F_J,
\end{equation}
This gives us the following ABHY polytope:
setting 
$S_I=-C_I$ for any non-singleton $I$, where $C_I$ are positive 
constants, and the polytope is given by the alternating sum
\begin{equation}\label{ABHY}
	F_I\geq 0,\quad -C_I=\sum_{J\neq \varnothing,J\subset I}(-1)^{|J|-|I|}F_J
\end{equation}
in the $\{F_I\}$-space.
By eq.\eqref{deffacet}, this polytope is the $n$-dimensional 
permutohedron $\mathscr P_n(\{C_I\})$, and it is nice that it admits a simple ABHY-like realization.
The ABHY construction of any generalized permutohedron can be obtained by 
setting $C_I=0$ in $\mathscr P_n(\{C_I\})$ for all $I$ not in its building set.

For example, the building set of $\mathscr A_2$ is $\mathbf B(\mathscr P_2)-\{\{0,2\}\}$,
its ABHY construction can be obtained from $\mathscr P_2$ by setting $C_{02}=0$, 
\[
	0=-C_{02}=F_{02}-F_2,
\]
therefore the whole ABHY conditions for $\mathscr A_2$ is 
\[
	-C_{01}=F_{01}-F_0-F_1,\quad -C_{12}=F_{12}-F_1-F_2,\quad 
	-C_{012}=-F_{01}-F_{12}+F_1,\quad F_I\geq 0,
\]
and its solution in $(F_1,F_2)$-space is 
\[
	\left\{\begin{aligned}
		F_1&\geq 0,F_2\geq 0\\
		F_0&=C_{01}+C_{12}+C_{012}-F_1-F_2\geq 0,\\
		F_{12}&=F_1+F_2-C_{12}\geq 0,\\
		F_{01}&=C_{012}+C_{12}-F_2\geq 0.
	\end{aligned}\right.
\]

For another example, consider the building set
\[
\mathbf{B}=\{\{0\},\{1\},\{2\},\{3\},\{0,1,2\},\{1,3\},\{0,1,2,3\}\}.
\]
It's a degeneration of $\mathscr P_3$, where
\[
C_{01},C_{02},C_{03},C_{12},C_{23},C_{013},C_{023},C_{123}=0,
\]
and its $u$ variables and $u$ equations can be found at the end of section (\ref{sect3.4}). By solving the ABHY conditions for these $C_I=0$, we get the vanishing facets
\begin{align*}
&F_{0i}=F_0+F_i,\quad F_{12}=F_1+F_2,\quad F_{23}=F_2+F3,\\
&F_{013}=F_{13}+F_0,\quad F_{023}=F_0+F_2+F_3,\quad F_{123}=F_2+F_{13},
\end{align*}
and the remaining ABHY conditions
\[
C_{13}=F_1+F_3-F_{13},\quad C_{012}=F_{0}+F_{1}+F_2-F_{012},
\quad C_{0123}=F_{012}+F_{13}-F_1,
\]
its solution in $(F_1,F_2,F_3)$-space is 
\[
	\left\{\begin{aligned}
		F_1&\geq 0,F_2\geq 0,F_3\geq 0\\
		F_0&=C_{012}+C_{0123}+C_{13}-F_1-F_2-F_3\geq 0,\\
		F_{13}&=F_1+F_3-C_{13}\geq 0,\\
		F_{012}&=C_{0123}+C_{13}-F_3\geq 0.
	\end{aligned}\right.
\]

Thanks to eq.\eqref{SinF}, using the picture of big polyhedron, it gives $u$ variables of permutohedron $\mathscr P_n$
\begin{equation}
  u_I=\prod_{J\supset I}x_J^{(-1)^{|I|-|J|}}. \label{uvarforPn}
\end{equation}
since
\[
	\prod_{I\subset [0,n]}u_I^{F_I}=
	\prod_{I\subset [0,n]}x_I^{S_I}=
	\prod_{ J\subset I\subset [0,n]}
	x_I^{(-1)^{|J|-|I|}F_J}=
	\prod_{J\subset [0,n]}
	\prod_{I\supset J}(x_I^{(-1)^{|J|-|I|}})^{F_J}.
\]
Let us look at the $n=2,3$ examples:

\paragraph*{2d permutohedron}
In this case, the $u$ variables can be written in terms of $x$ variables as:
\[
	u_{0}  = \frac{x_0 x_{012}}{x_{01}x_{02}},\quad 
	u_{1}  = \frac{x_1 x_{012}}{x_{01}x_{12}},\quad
	u_{2}  = \frac{x_2 x_{012}}{x_{02}x_{12}},\quad
	u_{01} = \frac{x_{01}}{x_{012}},\quad 
	u_{02} = \frac{x_{02}}{x_{012}},\quad
	u_{12} = \frac{x_{12}}{x_{012}}.
\]
The $u$ equations are:
\begin{align}\label{perm2}
	&1-u_{0}=u_{1} u_{2} u_{12}^2,
	1-u_{1}=u_{0} u_{2} u_{02}^2,
	1-u_{2}=u_{0} u_{1} u_{01}^2,\nonumber\\
	&1-u_{01}=u_{2} u_{02} u_{12},
	1-u_{02}=u_{1} u_{01} u_{12},
	1-u_{12}=u_{0} u_{01} u_{02}.
\end{align}
which are the same as the $u$ equations for $\mathscr{B}_{2}$.

\paragraph*{3d permutohedron}
In this case, 
the $u$ variables can be written in terms of $x$ variables as:
\begin{align*}
  u_{0} = \frac{x_{0}x_{012}x_{013}x_{023}}{x_{01}x_{02}x_{03}x_{0123}} \:,\qquad  
  u_{01} = \frac{x_{01}x_{0123}}{x_{012}x_{013}} \:,\qquad
  u_{02} = \frac{x_{02}x_{0123}}{x_{012}x_{023}}\:, \qquad 
  u_{012} = \frac{x_{012}}{x_{0123}},
\end{align*}
and their cyclic permutations. The $u$ equations are
\begin{align}\label{perm3}
  1-u_{0} &= u_1 u_{2} u_{3} u_{12}^2 u_{23}^2  u_{13}^2 u_{123}^3 \left(1+ u_{0}  u_{023} u_{013} u_{012} u_{03} u_{01}  u_{02} \right)\:,  \nonumber\\
  1-u_{0 1} &= u_{2} u_{3}  u_{23}^2 u_{12} u_{13} u_{123}^2 u_{023}^2 u_{02} u_{03} \:,\nonumber\\
  1-u_{02} &= u_{1}u_{3} u_{13}^{2}u_{12}u_{23} u_{123}^{2} u_{013}^{2}u_{01}u_{03} \:,\nonumber\\
 1-u_{012} &= u_3 u_{03}u_{13} u_{23} u_{0 1 3} u_{0 2 3} u_{123}  \:,
\end{align}
and their cyclic permutations.The $u$ variables and $u$ equations of $\mathscr P_4$ can be found in appendix (\ref{appendix:B}).

\subsection{Permutohedra as binary geometries}

In this subsection, we will show that the $u_{I}$'s for the permutohedron $\mathscr{P}_{n}$, which are defined by eq.\eqref{uvarforPn},
have the desired binary property. Recall that $u_{I}$ and $u_{J}$ are compatible if and only if $I \subset J$ or $J\subset I$, otherwise they are incompatible. We will see that as  $u_{I} \to 0$, all the incompatible $u_J \to 1$. Furthermore, we will show that the $u$ variables compatible with $u_{I}$ become the $u$ variables for the facet $\mathscr{P}_{\lvert I \rvert-1}\times \mathscr{P}_{n-\lvert I\rvert}$. 

A crucial observation here is that $u_{I}\to 0$ is equivalent to $x_{I}\to 0 $ according to \eqref{uvarforPn} since all $x_{i}$ are positive. We can approach this limit by replacing $x_{i}$ with $ \epsilon x_{i}$ for all $i\in I$ then taking $\epsilon \to 0$. Then the remaining task is just to consider the behaviour of the other $u_{J}$'s under this limit. 

Let us first show that the $u$ variables incompatible with $u_{I}$ become 1 as $u_{I}$ goes to 0. It is obvious that $u_{I}$ and $u_{J}$ are incompatible if either (1) $I\cap J=  \varnothing$  or (2) $I\cap J \neq \varnothing,I, J $. For the first case, let us denote $K=[0,n]-I-J$ as the complement set of $[0,n]$ with respect to $I\cup J$. Then we find the logarithm of $u_{J}$ can be written as 
\begin{equation}
     \log u_{J} = \sum_{\kappa\subset K,\psi\subset I} (-1)^{\lvert\kappa\rvert+\lvert \psi\rvert }\log (x_{J}+x_{ \kappa} +x_{ \psi} )\:. \label{1incompuvarforPn}
\end{equation}
As $x_{i}\to 0$ with $i\in I$, we have 
\begin{equation}
   \log u_{J} \to \sum_{\kappa\subset K} (-1)^{\lvert \kappa\rvert } \log (x_{J}+x_{ \kappa})\Biggl(\sum_{\psi\subset I}(-1)^{\lvert\psi\rvert}\binom{\lvert I \rvert }{\lvert \psi \rvert }\Biggr)=0 \,,\label{incompn1}
\end{equation}
where the binomial expansion of $(1-1)^{\lvert I \rvert}$ has been used. For the second case, the logarithm of $u_{J}$ can be written as 
\begin{equation}
   \log u_{J} = \sum_{\psi'\subset I',\kappa\subset K}(-1)^{\lvert \psi' \rvert+ \lvert \kappa\rvert}  \log (x_{J'}+ x_{(I\cap J)}+x_{ \psi'}+x_{\kappa })\,,
\end{equation}
where $I'=I-J$, $J'=J-I$ and $K=[0,n]-(I\cup J)$. Then a similar argument as in \eqref{incompn1} gives $\log u_{J}\to 0$ under the limit of all $x_{i}\to 0$ with $i\in I$.

Now let us consider the behaviour of compatible $u$ variables under this limit. For $J\subset I$, we again introduce $I'=I-J$ and $K=[0,n]-I$, then the logarithm of $u_{J}$ can written as 
\begin{equation}
     \log u_{J}=\sum_{\psi'\subset I', \kappa \subset K} (-1)^{\lvert \psi'\rvert +\lvert \kappa \rvert} \log (x_{J}+x_{\psi'}+x_{\kappa})\,.
\end{equation}
Next, we replace $x_{i}$ with $ \epsilon x_{i}$ for all $i\in I$ and take $\epsilon \to 0$, then 
\begin{align}
   \log u_{J} &\to \lim_{\epsilon \to 0} \sum_{\psi'\subset I', \kappa \subset K} (-1)^{\lvert \psi'\rvert +\lvert \kappa \rvert} \log (\epsilon x_{J}+ \epsilon x_{\psi'}+x_{\kappa}) \nonumber  \\
    &=\lim_{\epsilon\to 0} \biggl(\log \epsilon+\sum_{\kappa\subset K,\kappa \neq \varnothing}\log x_{\kappa}\biggr)\Biggr( \sum_{\psi'\subset I'}(-1)^{\lvert\psi'\rvert}\binom{\lvert I \rvert }{\lvert  \psi' \rvert }\Biggr)
    +\sum_{\psi'\subset I'} (-1)^{\lvert \psi'\rvert }\log (x_{J}+x_{\psi'}) \nonumber \\
    &=\sum_{\psi'\subset I'} (-1)^{\lvert \psi'\rvert }\log (x_{J}+x_{\psi'}) \,.\label{subP1}
\end{align}
Obviously, the expression in the last line of \eqref{subP1} is a $u$ variable for the permutohedron $\mathscr{P}_{\lvert I\rvert -1}$. For $J\supset I$, under the limit of $x_I\to 0$, the logarithm of $u_{J}$ simply becomes 
\begin{equation}
   \log u_{J} = \sum_{\kappa \subset K} (-1)^{\kappa} \log (x_{I}+x_{J-I}+x_{\kappa}) 
   \to  \sum_{\kappa \subset K} (-1)^{\kappa} \log (x_{J-I}+x_{\kappa})\, ,
\end{equation}
where $K=[0,n]-J$, which is a $u$ variable for the permutohderon $\mathscr{P}_{n-\lvert I\rvert}$. 

\subsubsection{The $u$ equations of ${\mathscr P_n}$}

For a permutohedron $\mathscr P_n$, we can directly use eq.\eqref{GPfacet} to express $x_I$ in terms of $\{u_I\}$ which gives us many algebraic equations for the $u$ variables:
\[
\prod_{I}x_I^{S_I}=\prod_{J}u_J^{F_J}=\prod_{J}\prod_{I\subset J}u_J^{S_I}=\prod_{I}\biggl(\prod_{J\supset I}u_J\bigg)^{S_I},
\]
so
\begin{equation}\label{xinu}
x_I=\prod_{J\supset I}u_J
\end{equation}
for any $I\in \mathscr P_n$, where we introduce a fake $u$ variable 
$u_{[0,n]}:=x_{[0,n]}$. We know some algebraic equations of $\{x_I\}$, 
for example, linear equations $x_I+x_J=x_{I\cup J}$ for $I\cap J=\varnothing$, 
so
\begin{equation*}
	\prod_{K\supset I}u_K+\prod_{K\supset J}u_K=\prod_{K\supset I\cup J}u_K
\end{equation*}
or equivalently
\begin{equation}
\prod_{I\subset K,K\not\supset I\cup J}u_K +
\prod_{J\subset K,K\not\supset I\cup J}u_K = 1.
\end{equation}
We can in fact use the above relations to find the $u$ equations for $\mathscr{P}_n$ recursively as follows:

For $I,J$ such that $|I|=n$ and $J=[0,n]-I = \{ i \}$ we have 
$1-u_{I}= \prod_{i \in J} u_{J}$
which is already an $u$ equation for $u_{I}$. For $|I|=n-1,~J=\{i\}$ and $\{k\}=[0,n]-(I\cup J)$ we have 
$1- u_{I}u_{I\cup\{k\}} = \prod_{{\{i\} \in J}\atop{J\neq I\cup\{i\}}} u_J $ by rewriting $1- u_{I}u_{I\cup\{k\}}  =1-u_{I}+u_{I}(1-u_{I\cup\{k\}})$ and using the above equations we get 
\bea
1-u_{I}=\prod_{{\{i\} \in J}\atop{J\neq I\cup\{i\}}} u_J - u_{I} \prod_{\{i\} \in K} u_K =\prod_{{\{i\} \in J}\atop{J\neq I\cup\{i\}}} u_J \left(1- u_{I}  u_{I\cup\{i\}}\right)
\eea
More generally we can find the $u$ equation for any $I$ recursively by starting from $|I|=n$. We can argue this inductively by assuming we know the $u$ equations for all $I$ such that $|I|\geq k$.

We can now look at all the sets $\{I_1,\dots,I_n\}$ that contain $|I|=k-1$ but not $I\cup J$ and arrange them as $|I|<|I_1|\leq |I_2|\leq \cdots \leq |I_n|$
\begin{align*}
1-\prod_{I\subset K,K\not\supset I\cup J}u_K &=1-u_{I}u_{I_1}\cdots u_{I_n}= 1-u_I+u_I(1-u_{I_1}\cdots u_{I_n})\nonumber \\
&=(1-u_I)+u_I(1-u_{I_1})+u_I u_{I_1}(1-u_{I_2})+\cdots+u_I\prod_{i=1}^{n-1}u_{I_{i}}(1-u_{I_n})
\end{align*}
Since $|I|< |I_i|$ for any $i$, we already know the $u$ equations for $I_i$ and we can find the equation for $u_I$.

Let us consider the $2 d$ example for which the relations are :
\[
1-u_1 u_{12} =u_0 u_{02}, ~1-u_2 u_{12} =u_0 u_{01},~1-u_{12}= u_0 u_{01} u_{02},~1- u_{01}= u_2 u_{01}u_{12},~1-u_{02}= u_1 u_{02} u_{12}
\]
and the $u$ equations
\begin{align}\label{npu}
1-u_1&=u_0 u_{02}(1-u_1 u_{01}), \quad 1-u_2 =u_0 u_{01}(1-u_2 u_{02}),\quad 1-u_{12}= u_0 u_{01} u_{02} \nonumber\\
1- u_{01}&= u_2 u_{01}u_{12},\qquad \qquad \:1-u_{02} = u_1 u_{02} u_{12} \quad\qquad \qquad \: 1-u_0=u_1 u_{12}(1-u_0 u_{01})
\end{align}
which is different non-perfect set of $u$ equations for ${\mathscr P}_2$ unlike eq.\eqref{perm2}!

A nice fact about the $u$ equations we obtain is that $1-u_I$ are all at most linear and multi-term in $u_I$ for any $n$.
However, they do not make the factorization manifest at the level of $u$ equations as we take some $u_I \rightarrow 0$ all incompatible $u_J \rightarrow 1$ but the $u$ equations corresponding to other $u$'s are trivially satisfied thereby not reflecting the facets of ${\mathscr P}_n$ (as can be seen by taking $u_{12} \rightarrow 0$ in eq.\eqref{npu}).

The $u$ equations for $\mathscr{P}_n$ similar to eqs.\eqref{perm2}, \eqref{perm3} which do make factorization manifest are complicated polynomials 
in the corresponding $u$'s. 
The most general $u$ equations can be conjectured to be:
\bea \label{ueqforpn}
1-u_{I} = (1+h_{I,n}(u)) \prod_{J} u^{J||I}_{J} 
\eea
where $J$ runs over all sets incompatible with $I$, $J||I$ is a positive integer, called the ``compatibility degree'', and $h_{I,n}(u)$ is polynomial 
in $u_I$ such that $h_{I,n}|_{u_I =0} = 0$ which makes $\mathscr{P}_n$ binary as we proved using a different method in the previous section. But we do not have a closed form expression for $h_{I,n}(u)$ for all $n$ and $I$. We list some of them along with the ${\mathscr P_4}$ example in appendix (\ref{appendix:B}). 




Although the polynomial $h_{I,n}$ is very complicated for general $I$ and $n$, we can still say something about its  ``compatibility degree'' $J||I$ in eq.\eqref{ueqforpn}.
Using eq.\eqref{uvarforPn}, 
\begin{equation}\label{oneminusui}
1-u_I=1-\prod_{J\supset I}x_J^{(-1)^{|J|-|I|}}=\frac{\prod_{|J|-|I| \text{ odd}; J\supset I}x_J-\prod_{|J|-|I| \text{ even}; J\supset I}x_J}{\prod_{|J|-|I| \text{ odd}; J\supset I}x_J}=:\frac{N_{I,n}(x)}{\prod_{|J|-|I| \text{ odd}; J\supset I}x_J},
\end{equation}
where $N_{I,n}(x)$ is a polynomial of degree $2^{|\bar I|-1}$ in $x$.
It's easy to see that $x_i$ for $i\not\in I$ is a factor of $N_{I,n}(x)$ by setting $x_i=0$, so we can define a new polynomial $f_{I,n}$ of degree $2^{|\bar I|-1}-|\bar I|$ with non-negative coefficients such that
\[
N_{I,n}(x)=f_{I,n}(x)\prod_{i\not\in I}x_i.
\]
Using eq.\eqref{xinu},
\[
\prod_{i\not\in I}x_i\bigg/\prod_{|J|-|I| \text{ odd}; J\supset I}x_J
=\prod_{J\not\supset I} u_J^{|J|-|I\cap J|}\prod_{J\supsetneq I} u_J^{|J|-|I|-2^{|J|-|I|-1}}.
\]
Since $|J|-|I|-2^{|J|-|I|-1}\leq 0$ for $J\supsetneq I$, it's natural to rewrite eq.\eqref{oneminusui} as
\[
1-u_I=g_{I,n}\prod_{J\not\supset I} u_J^{|J|-|I\cap J|},
\]
where 
\[
g_{I,n}=f_{I,n}\bigg /\prod_{J\supsetneq I} u_J^{2^{|J|-|I|-1}-(|J|-|I|)}
=f_{I,n}\bigg /\prod_{J\supset I, |J|-|I|\, \text{odd}\atop{|J|-|I|\geq 3}} x_J.
\]
It's easy to see that $g_{I,n}=1+h_{I,n}$ by taking $x_I\to 0$, where $h_{I,n}$ is a rational function of $x$ such that $h_{I,n}|_{x_I=0}=0$. In fact, 
\[
N_{I,n}|_{x_I=0}=\prod_{|J|-|I| \text{ odd}; J\supset I}x_{J-I}=\prod_{i\not\in I}x_i\prod_{J\supset I, |J|-|I|\, \text{odd}\atop{|J|-|I|\geq 3}} x_{J-I}=f_{I,n}|_{x_I=0}\prod_{i\not\in I}x_i\quad \Rightarrow\quad g_{I,n}|_{x_I=0}=1.
\]
Therefore, 
\begin{equation}\label{ueqofPn}
1-u_I=(1+h_{I,n})\prod_{J\not\supset I} u_J^{|J|-|I\cap J|},
\end{equation}
and we conjecture that $h_{I,n}$ is a polynomial of $u$ with non-negative coefficients, and the ``compatibility degree'' is 
\[
J||I=|J|-|I\cap J|\text{ for $I\not\subset J$},\quad  J||I=0 \text{ for $I\subset J$}.
\]
It's clear that any compatible $u_J$ for $J\subset I$ doesn't appear in the product because $I\cap J=J$.
Note that it's not symmetric when $|I|\neq |J|$.

\subsection{Generalized permutohedra as binary geometries}\label{sect3.4}

Inspired by the $u$ variables for permutohedra, we directly come up with the $u$ variables for generalized permutohedra. To this end, we first find all minimum extensions $I_{a}$ of $I$ in $\mathbf{B}$, that is there is no $J\in\mathbf{B}$ such that $I\subsetneq J\subsetneq I_{a}$, and we denote the collection of these extensions by $\mathscr{G}_{I}$. Then, the $u$ variables for the generalized permutohedron with the building set $\mathbf{B}$ are
\begin{equation}
   \log u_{I} :=  \log x_{I} +(-1)^{k}\sum _{\varnothing\neq\{I_{a_{1}},\dots,I_{a_{k}}\}\subset \mathscr{G}_{I}} \log x(I_{a_{1}},\ldots, I_{a_{k}})  \label{uvarforgenpermu}
\end{equation}
where the sum is over all nonempty subsets of $\mathscr{G}_{I}$, and we introduce
\[
x(I_{a_{1}},\ldots,I_{a_{k}}):=x_{I_{a_{1}}\cup \cdots \cup I_{a_{k}}}   
\]
to avoid too many subscripts. Obviously, an equivalent form of eq.\eqref{uvarforgenpermu} is
\begin{equation}
   \log u_{I} := (-1)^{k}\sum _{\{I_{a_{1}},\dots,I_{a_{k}}\}\subset \mathscr{G}_{I}} \log \bigl(x_{I}+x(I_{a_{1}}-I,\ldots, I_{a_{k}}-I)\bigr)  \label{uvarforgenpermu1}
\end{equation}
Let us consider several simple examples to illustrate this construction:
\begin{enumerate}[label=(\roman*)]
   \item For a $n$-dimensional hypercube, $\mathscr{G}_{\{0\}}=\{[0,1]\}$, $\mathscr{G}_{\{i\neq 0\}}=\{[0,i]\}$ and $\mathscr{G}_{[0,i]}=\{[0,i+1]\}$. This gives $u_{0}=x_{0}/x_{01}$, $u_{i}=x_{i}/x_{[0,i]}$ and $u_{[0,i]}=x_{[0,i]}/x_{[0,i+1]}$. These are indeed the $u$ variables for the $n$-dimensional hypercube since  $u_{i+1}+u_{[0,i]}=1$ for $i\in [0,n{-}1]$.
   \item For a $n$-dimensional associahedron,
   \[
   \mathscr{G}_{[i,j]}=\begin{cases}
      \{[i-1,j],[i,j+1]\}   & \quad \text{if $i\neq 0$ and $j\neq n $}\\
      \{[0,j+1]\} &\quad \text{if $i= 0$ and $j\neq n $} \\
      \{[i-1,j]\} &\quad \text{if $i\neq 0$ and $j= n $}
   \end{cases}   \:.
   \]
   This reproduces the $u$ variables \eqref{uvarforAn} for the $n$-dimensional associahedron.
   \item For a $n$-dimensional cyclohedron, $\mathscr{G}_{[i,i+k]}=\{[i-1,i+k],[i,i+k+1]\}$.  This reproduces the $u$ variables \eqref{uvarforBn} for the $n$-dimensional cyclohedron.
   \item For a $n$-dimensional permutohedron, $\mathscr{G}_{I}=\{I\cup\{j\}\vert j\in [0,n]-I \}$. This reproduces the $u$ variables \eqref{uvarforPn} for the $n$-dimensional permutohedron.
\end{enumerate}


For generalized permutohedra, as we have reviewed in sec.(\ref{GP}), $u_{I}$ and $u_{J}$ are incompatible if $I$ and $J$ are one of the following two cases:
\begin{enumerate}[label=(\arabic*)]
   \item $I\cap J=\varnothing$ and $I\cup J \in \mathbf{B}$ \:, 
   \item $I\cap J \neq \varnothing,I, J $ \:,
\end{enumerate}
while $u_{I}$ and $u_{J}$ are compatible if $I$ and $J$ satify one of the following three conditions:
\begin{enumerate}[resume,label=(\arabic*)]
   \item $J\subset I$\:,
   \item $I\subset J$ \:,
   \item $I\cap J=\varnothing$ and $I\cup J \notin \mathbf{B}$ \:.
\end{enumerate}
We will follow the same argument as in the case of permutohedra: replacing $x_{i}$ with $ \epsilon x_{i}$ for all $i\in I$ and taking $\epsilon \to 0$, then considering the behaviour of the other $u_{J}$'s under this limit case by case. 

For case (1), some elements in $\mathscr{G}_{J}$ are subsets of $I\cup J$ since $I\cup J \in \mathbf{B}$. We denote the collection of such elements by $\mathscr{G}^{I\cup J}_{J}$. Then the logarithm of $u_{J}$ can be written as 
\begin{equation}
   \log u_{J}= \sum_{\substack {\{J_{a_{1}},\ldots,J_{a_{k}}\}\subset \mathscr{G}^{I\cup J}_{J}\\ \{J_{b_{1}},\ldots J_{b_{l}}\}\subset \mathscr{G}_{J}-\mathscr{G}^{I\cup J}_{J}}}(-1)^{k+l}\log\Bigl( x_{J}+ x(J_{a_{1}}-J,\ldots,J_{a_{k}}-J, J_{b_{1}}-J,\ldots, J_{b_{l}}-J)\Bigr) \:. \label{logu1}
\end{equation}
Since the second term in eq.\eqref{logu1} goes to $x(J_{b_{1}}-J-I,\ldots, J_{b_{l}}-J-I)$ under the limits of $x_{i}\to 0$ with $i\in I$, $\log u_{J}$ goes to 0 under this limit as in eq.\eqref{1incompuvarforPn}. The whole argument can be carried over to case (2) by replacing $x_{J}$ with $x_{I\cap J}+ x_{J-I}$.

For case (3), we again denote $\mathscr{G}^{I}_{J}$ as the collection of the elements in $\mathscr{G}_{J}$ which are subsets of $ I$. After replacing $x_{i}$ with $\epsilon x_{i}$ for $i\in I$, we find the limit of $\log u_{J}$ as $\epsilon \to 0$ is
\begin{align}
  &\quad \lim_{\epsilon\to 0}  \sum_{\substack {\{J_{a_{1}},\ldots,J_{a_{k}}\}\subset \mathscr{G}^{I}_{J}\\ \{J_{b_{1}},\ldots J_{b_{l}}\}\subset \mathscr{G}_{J}-\mathscr{G}^{I}_{J}}}(-1)^{k+l}\log\Bigl( \epsilon x_{J}+\epsilon x(J_{a_{1}}{-}J,\ldots,J_{a_{k}}{-}J,(J_{b_{1}}{-}J)\cap I,\ldots, (J_{b_{l}}{-}J)\cap I) \nonumber \\
  & \qquad \qquad\qquad \qquad \qquad \qquad    + x (J_{b_{1}}{-}I,\ldots,J_{b_{l}}{-}I)\Bigr) \nonumber  \\
  &=\sum_{\{J_{a_{1}},\ldots,J_{a_{k}}\}\subset \mathscr{G}_{J}^{I}} (-1)^{k+l}\log\Bigl( x_{J}+ x(J_{a_{1}}{-}J,\ldots,J_{a_{k}}{-}J) \Bigr). 
\end{align}
Remarkably, these are $u$ variables for the generalized permutohedron with the building set $\mathbf{B}_{I}$ defined by \eqref{BI}.

Case (4) and (5) are quite trivial. For both cases, the limit behaviour of $u_{J}$ as $x_{I}\to 0$ can be simply obtained by replacing each $x_{K}$ in eq.\eqref{uvarforgenpermu} with $x_{K-I}$. 
Remarkably, these are the $u$ variables for the generalized permutohedron with the building set $ \mathbf{B}_{\bar{I}}$ defined by \eqref{BIbar}.

Here we just showed that binary geometries exist for generalized permutohedra and gave the corresponding $u$ variables. The $u$ equations are quite complicated in general, and we only have closed formulas for $\mathscr{A}_{n}$, $\mathscr{B}_{n}$ and their several degenerations. For permutohedra, we have just conjectured the $u$ equations are of form \eqref{ueqforpn}. More generally, the $u$ equations don't even take the form of \eqref{ueqforpn}. For example, consider the generalized permutohedron with the building set 
\[
\mathbf{B}=\{\{0\},\{1\},\{2\},\{3\},\{0,1,2\},\{1,3\},\{0,1,2,3\}\},
\]
the corresponding $u$ variables are
\begin{gather*}
   u_{0}=\frac{x_{0}}{x_{012}} \:,\quad  u_{1}=\frac{x_{1}x_{0123}}{x_{012}x_{13}} \:,\quad 
   u_{2}=\frac{x_{2}}{x_{012}} \:, \quad u_{3}=\frac{x_{3}}{x_{13}} \\
   u_{012}=\frac{x_{012}}{x_{0123}}\:,\qquad u_{13}=\frac{x_{13}}{x_{0123}} \:. \quad 
\end{gather*}
The $u$ equations are\footnote{If we introduce $\tilde{u}_{02}=u_{0}+u_{2}$, then $\tilde{u}_{02}$ together with the remaining 4 $u$'s form $\mathscr{A}_{2}$.}
\begin{gather*}
u_{0}+u_{2}+u_{1}u_{13}=1\:, \quad  u_{1}+(u_{0}+u_{2})u_{3}=1 \:, \quad u_{3}+u_{1}u_{012}=1\:, \\
u_{012}+u_{3}u_{13}=1\:, \quad u_{13}+(u_{0}+u_{2})u_{012}=1\:.
\end{gather*}
We leave the study of compatible degrees and $u$ equations for general permutohedra to future work.

\section{Discussions}

In this paper we introduce a large class of integrals,  ${\cal I}_{\bf B}$, which can be regarded as a family of rigid stringy canonical forms 
for generalized permutohedra ${\mathscr P}(\bf B)$, and we show that corresponding configuration spaces are always binary geometries. These rigid integrals satisfy the remarkable property that, at any ``massless'' pole, {\it i.e.} poles of the leading order $\Omega({\mathscr P}(\bf B) )$, the residue is given by a product of two such integrals, ${\cal I}_{{\bf B}_L} \times {\cal I}_{{\bf B}_R}$. Our results greatly extend the scope of binary geometries and associated integrals with such ``perfect'' factorization properties, which were originally thought to tie to the more special class of generalized associahedra for finite-type cluster algebras. For infinite families of cases as degenerations of associahedra and cyclohedra (which belong to both generalized permutohedra and generalized associahedra), we find additionally that the $u$ equations are ``perfect'' just like the case for cluster configuration spaces. Our preliminary investigations have opened up various new avenues to be explored, and we end with listing some open questions.

\paragraph{Generalized permutohedra vs. generalized associahedra} Our results provide an infinite number of examples generalizing ordinary string integrals (and cluster string integrals of type $B_n$), where the configuration space is determined by {\it linear} factors, or equivalently hyperplane arrangement. In this sense, it is fair to say that we have only explored a very special corner in the world of binary configuration spaces and associated stringy integrals. Even for cluster cases, it is unknown if the cluster configuration space of type $C_n$ can be realized by hyperplane arrangement (see also \cite{Li:2018mnq}), and for type $ D_n$ and exceptional cases, the conjecture is that the spaces cannot be obtained this way~\cite{Arkani-Hamed:2019plo}; we expect that these cases do not belong to integrals with ``linear'' factors. It would be highly desirable to look for a wider class of binary geometries which naturally incorporate both generalized permutohedra and generalized associahedra. 

On the other hand, it would be interesting to study the configuration spaces for generalized permutohedra in a way similar to the cluster case. The $u$ equations for {\it e.g.} ${\mathscr P}_d$ are certainly worth further investigations, and one could ask if there might be any analog/generalization of cluster algebra behind it. Another question concerns going beyond the positive part of the space since it is binary even in the complex case, and it would be nice to study complex space and connected components of the real space. For the cluster case, connected components are in bijection with sign patterns of the $u$ variables allowed by the so-called extended $u$ equations~\cite{Arkani-Hamed:2019plo}. It would be highly desirable to see if one could obtain all the extended $u$ equations for {\it e.g.} ${\mathscr P}_d$, which can help us understand these questions better. Relatedly, the number of connected components and other interesting properties of the space can be inferred by counting the number of points in the space over a finite field; since we are dealing with hyperplane arrangement, the point count $N(p)$ is always a polynomial. Just to illustrate the counting with some examples, since ${\mathscr P}_2={\mathscr B}_2$, the point count is $N(p)=(p-3)^2$, and if we plug in $p=-1$ we have $|N(-1)|=16$ connected components; we find for ${\mathscr P}_3$, $N(p)=(p-4)(p^2-10p+26)$, thus there are $185$ connected components, and for ${\mathscr P}_4$, $N(p)=p^4- 30 p^3+ 345 p^2- 1785 p+3485$ thus $5646$ connected components. 

Moreover, since we have found a variety of such geometries with perfect $u$ equations from degenerating ${\mathscr A}_n$ and ${\mathscr B}_n$, it'd be very interesting to explore more cases with perfect $u$ equations from degenerations of other types, {\it e.g.} ${\mathscr C_n}$ and ${\mathscr D_n}$. Along this direction, an intriguing possibility is to classify binary geometries with perfect $u$ equations from degenerating cluster cases; since any product of such geometries still has perfect $u$ equations, this amounts to find all elementary, non-factorizing geometries with perfect $u$ equations. More importantly,  what are the possible interpretations of these degenerations from cluster algebra; even for ${\mathscr A}_{n-3}$, what is the meaning of these degenerations of ${\cal M}_{0,n}$ and the string integrals?

\paragraph{Generalized string integrals and applications} One of the most pressing questions is to study further these stringy integrals and their potential applications to physics. We would like to understand why these integrals factorize at finite $\alpha'$ just as cluster stringy integrals, though most of them are not related to cluster algebra. Recall that cluster stringy integrals of classical types are natural $\alpha'$-deformation of bi-adjoint $\phi^3$ amplitudes through one-loop~\cite{Arkani-Hamed:2019vag,Salvatori:2018aha}, and in general the physical meaning of these integrals and their leading orders are unclear. A special class of generalized permutohedra, called Cayley polytopes, have natural interpretations as the sum of certain tree-level $\phi^3$ diagrams~\cite{Gao:2017dek,He:2018pue}; for these cases, the stringy integrals provide a rigid $\alpha'$-deformation of these $\phi^3$ ``amplitudes''~\footnote{For other applications of generalized permutohedra to physics, see~\cite{early2017generalized,Early:2018zuw}.}. It would be interesting to see what role do generalize permutohedra play in the general story of formulating scattering amplitudes as differential forms~\cite{Arkani-Hamed:2017tmz, He:2018okq, Arkani-Hamed:2017mur, Arkani-Hamed:2019vag}.

We have not touched various properties of these stringy integrals, such as recurrence relations, complex (and real) integrals with two different ``orderings'' and scattering (saddle-point) equations and pushforward. It would be interesting to understand these points better, especially for the ${\mathscr P}_d$ case, which is completely symmetric in the labels. The scattering equations read
\bea
\frac{S_i}{x_i}=\sum_{I \ni i} \frac{C_I}{x_I}\,,
\eea
which provide a one-to-one map from ${\bf x}\in \mathbb{R}_{\geq 0}^d$ to ${\mathscr P}_d$, and $\Omega({\mathscr P}_d)$ is given by the pushforward via summing over the solutions. By plugging in $p=1$ in the point count $N(p)$, one can obtain the number of solutions (saddle points), which also gives the dimension of the space of integral functions~\cite{Arkani-Hamed:2019mrd}. For example, for $d=2, 3, 4$, we have $4, 51, 2016$ solutions respectively. 

One class of examples which are particularly interesting are the Stokes polytopes and Accordiohedra~\cite{Banerjee:2018tun,Raman:2019utu} associated with planar $\phi^{p}$ amplitudes. A nice fact about Stokes polytopes~\cite{Chapoton, Raman:2019utu} is that any product of associahedra ${\mathscr A}_{i_1}\times \cdots \times {\mathscr A}_{i_k}$ is a Stokes polytope of dimension $n=i_1+\cdots+i_k$; more interestingly, degenerations of ${\mathscr A_3}$ with perfect $u$ equations correspond to $3d$ Stokes polytopes. Furthermore every Stokes polytope is a Minkowski sum of associahedra \cite{Baryshnikov} and thus it would be natural to consider them because these might provide a nice class of physically relevant examples which also play a role in classifying binary geometries with perfect $u$ equations.

\paragraph {Beyond linear factors}
Throughout the paper we have considered stringy integrals with linear factors whose Newton polytopes are coordinate simplices. In fact, one can discover more polytopes which also have perfect $u$ equations, once we consider non-linear factors. One particularly simple class of examples we discover concern polytopes that are Minkowski sum of $n$ line intervals and $n-1$ triangles, with Newton polynomials:  
\bea \label{new}
\prod_{i=1}^{n} (1+x_i )^{-\alpha' c_i}  \prod_{i=1}^{n-1} (1+x_i +x_i x_{i+1})^{-\alpha' c_{i, i+1}}\,.
\eea 
It turns out this gives a family of simple $n$-dimensional polytopes with $3n-1$ facets and the number of vertices is given by the Pell number $P_n$ (recursively defined as $ P_n=2 P_{n-1} +P_{n-2}$ with $P_1=1,~ P_2=2$). We refer to the polytope as $\mathscr{X}_n$.

The big polyhedron is a simplex, and we find that the $u$ variables take a very simple form:
\bea
u_1 &=& \frac{p_n}{q_n}, ~u_2 = \frac{p_{n-1}q_n}{p_{n-1,n}},~u_3 = \frac{p_{n-2}q_{n-1}}{p_{n-2,n-1}}, \dots,
u_{n-2} = \frac{p_1 q_2}{p_{12}}, ~u_{n-1} = \frac{p_{2}q_3}{p_{23}},~u_n = \frac{p_{3}q_{4}}{p_{34}}, \nonumber \\
u_{n+1} &=& \frac{1}{q_1}, ~u_{n+2} = \frac{p_{12}}{q_{1}q_{2}},~u_{n+3} = \frac{q_{1}}{p_{23}}, \dots,
u_{3n-1} = \frac{ q_{n-1}}{p_{n-1n}} ,
\eea
where we denote $p_i :=x_i$, $q_i := 1+x_i$ and $p_{i~i+1}:=1+x_i+ x_i x_{i+1}$. Remarkably, for any $n$, we have perfect $u$ equations which fall into three types viz. $1-u$ being the product of two, three or four $u$'s. The first type has 4 equations:
\[
1-u_{1} =u_{3n-2} u_{3n-1},~
1-u_{n-2} =u_{n+1} u_{n+3},~
1-u_{n+1} =u_{n-2} u_{n+2},~
1-u_{3n-1} = \begin{cases}
u_{2} u_{3} , n=3\\
u_{1} u_{4} , n=4\\
u_{1} u_{2} , n\geq 5
\end{cases},
\]
with the corresponding facets being $\mathscr{X}_{n-1}$. The second type has $n-3$ equations
\[
1-u_{n+4} =u_{n} u_{n+2}u_{n+4} u_{n+6},
1-u_{n+6+2 i} =u_{n-3-i} u_{n+4+2 i} u_{n+5+2 i} u_{n+8+2 i}~ {\rm for}~i=0,\dots,n-5;
\]
with the corresponding facets being $\mathscr{A}^{m}_1 \times \mathscr{X}_{n-m-1}$. The third type has $2n-2$ equations, but we do not yet have a complete classification of these facets. It would be nice to see if $\mathscr{X}_n$'s are equivalent to some degenerations of $\mathscr{P}_n$, or if they are genuinely new binary geometries with perfect $u$ equations.  
%
%
%
%

\paragraph{Stringy integrals at finite $\alpha'$}

An important open question for stringy canonical form in general is to understand structures both at higher orders in the $\alpha'$-expansion and even at finite $\alpha'$. 
The leading order of \eqref{stringycanonicalform} computes the canonical form of the corresponding polytope, which does not depend on the coefficients of the polynomials, but higher orders do, and it would be interesting to study $\alpha'$ expansion for stringy integrals for {\it e.g.} permutohedra. 
 
Moreover, having closed form expressions at finite $\alpha'$ is certainly desirable. Our preliminary study already provides some examples, and let us look at a toy example which was considered in~\cite{Arkani-Hamed:2019mrd}:
 \bea
I(\alpha') &=& \int_{0}^{\infty} \int_{0}^{\infty} \frac{\mathrm{d}x}{x} ~\frac{\mathrm{d}y}{y} \left( \frac{1}{x} + \frac{1}{y}+ x y \right)^{-\alpha' }. \nonumber 
\eea 
We can use the symmetry in $x,~y$ to make a symmetric change of variables to $u=x+y$ and $v =x y$, and the integral can be easily evaluated as
 \bea
I(\alpha') &=& \int_{0}^{\infty} \int_{0}^{\frac{u^2}{4}} \frac{2 \mathrm{d}u ~ \mathrm{d}v}{v \sqrt{u^2-4 v} }\left( \frac{u}{v} + v  \right)^{-\alpha'}= \frac{\Gamma{\left[\frac{\alpha'}{3}\right]^{3}}}{3 \Gamma{ \left[\alpha'\right]}}. \nonumber
\eea
%

This is in fact a special class of a large class of integrals which all correspond to the simplex case. By using the integral identity 
\[\int_{{\mathbb R_{+}^{n}}} \prod_{i=1}^{n} \frac{\mathrm{d} x_i }{x_i}~x_i^{a_i}  f\left(\sum_{i=1}^{n} x_i \right)   = \frac{\prod_{i=1}^n \Gamma{(a_i)}}{\Gamma{(\sum_{i=1}^n a_i)}} \int_{0}^{\infty} f(\beta) \beta^{\sum_{i=1}^{n} a_n -1} \mathrm{d}\beta,
\]
we can easily evaluate the following general integral for simplex:
\bea
I(\alpha',~X_i,~c ) &=&  \int_{0}^{\infty} \cdots \int_{0}^{\infty} \prod_{i=1}^{n} ~\mathrm{d}x_i ~x_i^{\alpha' X_i -1}  \left( b+\sum_{i=1}^n a_i x_i^{m_i} \right)^{-\alpha' c } \nonumber  \\
&=& \frac{b^{\alpha' \left(\sum_{i=1}^n \frac{X_i}{m_i}-c \right) }}{(\prod_{i=1}^n m_i) (\prod_{i=1}^n a_i^{\alpha' X_i /m_i})} \int_{0}^{\infty} \cdots \int_{0}^{\infty} \prod_{i=1}^{n} ~\mathrm{d}y_i ~y_i^{ \frac{\alpha' X_i}{m_i} -1}  \left( 1+\sum_{i=1}^n y_i \right)^{-\alpha' c } \nonumber  \\
&=&\frac{b^{\alpha' \left(\sum_{i=1}^n \frac{X_i}{m_i}-c \right) }}{(\prod_{i=1}^n m_i) (\prod_{i=1}^n a_i^{\alpha' X_i /m_i})} B\left(\frac{ \alpha' X_1}{m_1},\dots,~\frac{ \alpha' X_n}{m_n}, ~ \alpha' \left(c- \sum_{i=1}^n \frac{X_i}{m_i} \right) \right). \nonumber \eea
All stringy integrals (including the ones with non-linear factors) can in principle be evaluated as linear combination of A-hypergeometric functions with special arguments by identification the with a GKZ system \cite{saito2013grobner,gelfand1990generalized} (see also \cite{delaCruz:2019skx}). We shall leave the evaluation of more stringy integrals and related topics to future work.

\section*{Acknowledgements}
We are grateful to Nima Arkani-Hamed, Thomas Lam, Hugh Thomas for stimulating discussions and collaborations on related projects. P.R. would like to thank the generous hospitality of the Institute of Theoretical Physics, Chinese Academy of Sciences, Beijing during a visit when part of this work was done. The research of S.H., Z.L. and C.Z. is supported in part by NSF of China under Grant No. 11947302 and 11935013. 

\appendix 

\section{Appendix: Degenerations of ${\mathscr P_n}$ which lead to products}\label{appendix:A}
In this section we consider degenerations of $\mathscr{P_n}$ give the most general products. 

The maximal set of a building set ${\mathbf B_{max}}$ is the set of all inclusion maximal elements of the building set ${\mathbf B}$. A building set ${\mathbf B}$ is connected if ${\mathbf B_{max}}$ has a single element\cite{Postnikov:2006}.

\vspace{1ex}

\noindent {\bf Theorem:} If ${\mathbf B_1},{\mathbf B_2},\dots {\mathbf B_n}$ are connected subsets of a building set ${\mathbf B}$ then the corresponding nestohedron ${\mathscr N_{{\mathbf B}}}$ is isomorphic to the product of the nestohedra ${\mathscr N}_{ {\mathbf B_1}} \times{\mathscr N}_{ {\mathbf B_2}} \cdots \times {\mathscr N}_{{\mathbf B_n}}$\cite{Postnikov:2006}.

\vspace{1ex}

The result follows as for any pair of connected building sets ${\mathbf B_{i}},{\mathbf B_{j}}$ either  ${\mathbf B_{i}} \cup {\mathbf B_{j}} = \varnothing $ in which case ${\cal I}_{{\mathbf B_i}\cup {\mathbf B_j}} = {\cal I}_{\mathbf B_i} \times {\cal I}_{\mathbf B_j}$ or ${\mathbf B_{i}} \cup {\mathbf B_{j}} = \{k\}$ in this we can rescale the $x_l \rightarrow x_k x_l$ for $l \in {\mathbf B_{j}} $ to decouple the integrals. Thus we always get ${\cal I}_{\mathbf B}={\cal I}_{\mathbf B_1}\times{\cal I}_{\mathbf B_2}\cdots \times{\cal I}_{\mathbf B_n}$. \\
Using the above result we construct a degeneration of  ${\mathscr P_n}$ which are the most general products {\it i.e.}, $\prod_{p_i} X_{p_i}$ such that $\sum^{k}_{i=1} p_{i} = n$ by dividing the $(n+1)$ singlets into $k$ parts 
\[
[0,p_1],\quad [p_1,p_1+p_2],\quad \dots ,\quad [\sum^{k-1}_{i=1}p_{i},\sum^{k}_{i=1}p_{i}= n]
\]
and using them to construct the building sets $\mathbf B_{p_i}$ of $X_{p_i}\left[\sum^{(i-1)}_{j=1}p_{j},\sum^{i}_{j=1}p_{j}\right].$ 

{\bf Example:} degenerations of ${\mathscr P_4}$ into ${\mathscr A_1^{4}}$, ${\mathscr A_1^2} \times{\mathscr A_2}$, ${\mathscr A_1} \times {\mathscr A_3}$, ${\mathscr A^{2}_2}$, ${\mathscr A_1^2} \times {\mathscr B_2}$, ${\mathscr A_1} \times {\mathscr B_3}$,${\mathscr A_1} \times {\mathscr P_3}$,$ {\mathscr A_2}\times {\mathscr B_2}$ and ${\mathscr B^{2}_2}$.
 
For ${\mathscr A_1^{4}}$ the building set is $ \{ {0},{1},{2},{3},{4},{01},{12},{23},{34} \}$ where $\{0,01\} , \{1,12\},\{2,23\},\{3,34\}$ generate the 4 ${\mathscr A_1}$'s which we denote as ${\mathscr A_1}(0,1) \times {\mathscr A_1}(1,2) \times {\mathscr A_1}(2,3) \times {\mathscr A_1}(3,4) $.

For ${\mathscr A_1}^{2} \times {\mathscr A_2}$ the building set is $ \{ {0},{1},{2},{3},{4},{01},{12},{23},{34},{234} \}$ generated by ${\mathscr A_1}(0,1)\times {\mathscr A_1}(1,2) \times {\mathscr A_2}(2,3,4)$'s.

For  ${\mathscr A_1} \times {\mathscr A_3}$ the building set is $ \{ {0},{1},{2},{3},{4},{01},{12},{23},{34},{123},{234},{1234} \}$ generated by  ${\mathscr A_1}(0,1)\times {\mathscr A_3}(1,2,3,4)$.

For  ${\mathscr A_2^{2}}$ the building set is $ \{ {0},{1},{2},{3},{4},{01},{12},{23},{34} ,{012},{234}\}$ generated by ${\mathscr A_2}(0,1,2)\times {\mathscr A_2}(2,3,4)$

For ${\mathscr A_1}^{2} \times {\mathscr B_2}$ the building set is $ \{ {0},{1},{2},{3},{4},{01},{12},{23},{34},{24},{234} \}$ generated by ${\mathscr A_1}(0,1)\times {\mathscr A_1}(1,2) \times {\mathscr B_2}(2,3,4)$'s.

For ${\mathscr A_1} \times {\mathscr B_3}$ the building set is $ \{ {0},{1},{2},{3},{4},{01},{12},{23},{34},{14},{123},{234},{134},{124},{1234} \}$ generated by  ${\mathscr A_1}(0,1)\times {\mathscr B_3}(1,2,3,4)$.

For ${\mathscr A_1} \times {\mathscr P_3}$ the building set is $ \{ {0},{1},{2},{3},{4},{01},{12},{23},{34},{14},{13},{24},{123},{234},{134},{124},{1234} \}$ generated by  ${\mathscr A_1}(0,1)\times {\mathscr P_3}(1,2,3,4)$.

For ${\mathscr A_2} \times {\mathscr B_2}$ the building set is $ \{ {0},{1},{2},{3},{4},{01},{12},{23},{34},{24},{012},{234}\}$ generated by ${\mathscr A_2}(0,1,2)\times {\mathscr B_2}(2,3,4)$

For ${\mathscr B_2^{2}}$ the building set is $ \{ {0},{1},{2},{3},{4},{01},{12},{02},{23},{34},{24} ,{012},{234}\}$ generated by ${\mathscr B_2}(0,1,2)\times {\mathscr B_2}(2,3,4)$.

We can similarly find a degeneration of ${\mathscr P_n}$ which would correspond to any product.

\section{Appendix: Some $u$ equations of ${\mathscr P_n}$}\label{appendix:B}

Some of the $u$ equations for the $n$-dimensional permutohedron are:\\
For $|I|=n$ or $n-1$ we have  
\[
1-u_{I} = \prod_{J\not\supset I} u_{J}^{|J|-|I\cap J|}
\]
and for $|I| =n-2$,
\[
1-u_{I}= \left(1+\prod_{J\supset I}u_J\right)\prod_{J\not\supset I} u_{J}^{|J|-|I\cap J|}
\]
The other $u$ equations are not two or three-term equations,
and in general it has the form 
\[
1-u_I=(1+h_{I,n}(u))\prod_{J\not\supset I} u_{J}^{|J|-|I\cap J|},
\]
where $h_{I,n}(u)$ is conjectured to be a polynomial of $\{u_J\,|\, J\supset I\}$ such that $h_{I,n}(u)|_{u_I=0}=0$.

Let us write down the ${\mathscr P_4}$ example:
\[
u_0= \frac{x_{0} x_{012} x_{013} x_{014} x_{023} x_{024} x_{034} x_{01234}}{x_{01} x_{02} x_{03} x_{04} x_{0123} x_{0124} x_{0134} x_{0234}},
u_{01} =  \frac{x_{01} x_{0123} x_{0124} x_{0134}}{x_{012} x_{013} x_{014} x_{01234}},
u_{012} = \frac{x_{012} x_{01234}}{x_{0123} x_{0124}},
u_{0123} = \frac{x_{0123}}{x_{01234}}
\]
The $u$ equations are 
\begin{align*}
   &1-u_0=u_{1} u_{2} u_{3} u_{4} u_{12}^2 u_{13}^2 u_{14}^2 u_{23}^2 u_{24}^2 u_{34}^2 u_{123}^3 u_{124}^3 u_{134}^3 u_{234}^3u_{1234}^4 
    (1+h_{\{0\},4}),\\
   &1-u_{01}=u_{2} u_{3} u_{4} u_{02} u_{03} u_{04} u_{12} u_{13} u_{14} u_{23}^2 u_{24}^2 u_{34}^2 u_{023}^2 u_{024}^2 u_{034}^2 u_{123}^2 u_{124}^2 u_{134}^2 u_{234}^3 u_{0234}^3 u_{1234}^3(1+h_{\{0,1\},4}),\\
   &1-u_{012}=u_{3} u_{4} u_{03} u_{04} u_{13} u_{14} u_{23} u_{24} u_{34}^2 u_{013} u_{014} u_{023} u_{024} u_{034}^2 u_{123} u_{124} u_{134}^2 u_{234}^2 u_{0134}^2 u_{0234}^2 u_{1234}^2,\\ 
   &1-u_{0123}=u_{4} u_{04} u_{14} u_{24} u_{34} u_{014} u_{024} u_{034} u_{124} u_{134} u_{234} u_{0124} u_{0134} u_{0234} u_{1234}
\end{align*}
where 
\[
h_{\{0,1\},4}=u_{01} u_{012} u_{013} u_{014} u_{0123} u_{0124} u_{0134},
\]
\begin{align*}
h_{\{0\},4}(u)=&\,u_{0} u_{01}^3 u_{02} u_{03} u_{04} u_{012}^3 u_{013}^3 u_{023} u_{024} u_{034} u_{0123}^2 u_{0124}^2 u_{0134}^2 u_{0234} u_{014}^3\\
&+u_{0} u_{01} u_{02} u_{03} u_{04} u_{012}^2 u_{013}^2 u_{023} u_{024} u_{034} u_{0123}^2 u_{0124}^2 u_{0134}^2 u_{014}^2\\
&+u_{0}^2 u_{01}^2 u_{02}^2 u_{03}^2 u_{04}^2 u_{012}^2 u_{013}^2 u_{023}^2 u_{024}^2 u_{034}^2 u_{0123} u_{0124} u_{0134} u_{0234}^2 u_{014}^2\\
&+u_{0} u_{01}^2 u_{02} u_{03} u_{04} u_{012}^2 u_{013}^2 u_{023} u_{024} u_{034} u_{0123} u_{0124} u_{0134} u_{0234}^2 u_{014}^2\\
&+u_{0} u_{01} u_{02} u_{03} u_{04} u_{012} u_{013} u_{023} u_{024} u_{034} u_{0234} u_{014}.
\end{align*}
The other $u$ variables and $u$ equations can be obtained by permutation of the indices. 

The 10 facets of $\mathscr P_4$ obtained by setting $u_{i}, u_{i j k l} \rightarrow 0$ are all $\mathscr{P}_3$ and the 20 facets obtained by setting $u_{i j},~u_{i j k} \rightarrow 0$ are all $\mathscr{A}_1 \times \mathscr{P}_2$ . 

\bibliographystyle{utphys}
\bibliography{draft_ABP}

\providecommand{\href}[2]{#2}\begingroup\raggedright\begin{thebibliography}{10}

\bibitem{Arkani-Hamed:2019mrd}
N.~Arkani-Hamed, S.~He, and T.~Lam, ``{Stringy Canonical Forms},''
\href{http://arxiv.org/abs/1912.08707}{{\ttfamily arXiv:1912.08707 [hep-th]}}.

\bibitem{Mellin}
L.~Nilsson and M.~Passare, ``Mellin transforms of multivariate rational
  functions,'' \href{http://dx.doi.org/10.1007/s12220-011-9235-7}{{\em Journal
  of Geometric Analysis} {\bfseries 23} (10, 2010) }.

\bibitem{Eulermellin}
C.~Berkesch, J.~Forsg{\aa}rd, and M.~Passare, ``Euler-mellin integrals and
  a-hypergeometric functions,''
  \href{http://dx.doi.org/10.1307/mmj/1395234361}{{\em Michigan Mathematical
  Journal} {\bfseries 63} no.~1, (Mar., 2014) 101--123}.

\bibitem{Panzer:2019yxl}
E.~Panzer, ``{Hepp's bound for Feynman graphs and matroids},''
  \href{http://arxiv.org/abs/1908.09820}{{\ttfamily arXiv:1908.09820
  [math-ph]}}.

\bibitem{Brown:2009qja}
F.~C.~S. Brown, ``{Multiple zeta values and periods of moduli spaces M 0 ,n ( R
  )},'' \href{http://dx.doi.org/10.24033/asens.2099}{{\em Annales Sci. Ecole
  Norm. Sup.} {\bfseries 42} (2009) 371},
\href{http://arxiv.org/abs/math/0606419}{{\ttfamily arXiv:math/0606419
  [math.AG]}}.

\bibitem{aomoto2011theory}
K.~Aomoto, M.~Kita, T.~Kohno, and K.~Iohara, {\em Theory of hypergeometric
  functions}.
\newblock Springer, 2011.

\bibitem{Mizera:2017rqa}
S.~Mizera, ``{Scattering Amplitudes from Intersection Theory},''
  \href{http://dx.doi.org/10.1103/PhysRevLett.120.141602}{{\em Phys. Rev.
  Lett.} {\bfseries 120} no.~14, (2018) 141602},
\href{http://arxiv.org/abs/1711.00469}{{\ttfamily arXiv:1711.00469 [hep-th]}}.

\bibitem{Mastrolia:2018uzb}
P.~Mastrolia and S.~Mizera, ``{Feynman Integrals and Intersection Theory},''
  \href{http://dx.doi.org/10.1007/JHEP02(2019)139}{{\em JHEP} {\bfseries 02}
  (2019) 139}, \href{http://arxiv.org/abs/1810.03818}{{\ttfamily
  arXiv:1810.03818 [hep-th]}}.

\bibitem{Mizera:2019vvs}
S.~Mizera and A.~Pokraka, ``{From Infinity to Four Dimensions: Higher Residue
  Pairings and Feynman Integrals},''
  \href{http://dx.doi.org/10.1007/JHEP02(2020)159}{{\em JHEP} {\bfseries 02}
  (2020) 159}, \href{http://arxiv.org/abs/1910.11852}{{\ttfamily
  arXiv:1910.11852 [hep-th]}}.

\bibitem{Brown:2018omk}
F.~Brown and C.~Dupont, ``{Single-valued integration and superstring amplitudes
  in genus zero},''
\href{http://arxiv.org/abs/1810.07682}{{\ttfamily arXiv:1810.07682 [math.NT]}}.

\bibitem{Arkani-Hamed:2017mur}
N.~Arkani-Hamed, Y.~Bai, S.~He, and G.~Yan, ``{Scattering Forms and the
  Positive Geometry of Kinematics, Color and the Worldsheet},''
  \href{http://dx.doi.org/10.1007/JHEP05(2018)096}{{\em JHEP} {\bfseries 05}
  (2018) 096}, \href{http://arxiv.org/abs/1711.09102}{{\ttfamily
  arXiv:1711.09102 [hep-th]}}.

\bibitem{Cachazo:2013iea}
F.~Cachazo, S.~He, and E.~Y. Yuan, ``{Scattering of Massless Particles:
  Scalars, Gluons and Gravitons},''
  \href{http://dx.doi.org/10.1007/JHEP07(2014)033}{{\em JHEP} {\bfseries 07}
  (2014) 033}, \href{http://arxiv.org/abs/1309.0885}{{\ttfamily arXiv:1309.0885
  [hep-th]}}.

\bibitem{Arkani-Hamed:2017tmz}
N.~Arkani-Hamed, Y.~Bai, and T.~Lam, ``{Positive Geometries and Canonical
  Forms},'' \href{http://dx.doi.org/10.1007/JHEP11(2017)039}{{\em JHEP}
  {\bfseries 11} (2017) 039},
\href{http://arxiv.org/abs/1703.04541}{{\ttfamily arXiv:1703.04541 [hep-th]}}.

\bibitem{Cachazo:2013hca}
F.~Cachazo, S.~He, and E.~Y. Yuan, ``{Scattering of Massless Particles in
  Arbitrary Dimensions},''
  \href{http://dx.doi.org/10.1103/PhysRevLett.113.171601}{{\em Phys. Rev.
  Lett.} {\bfseries 113} no.~17, (2014) 171601},
  \href{http://arxiv.org/abs/1307.2199}{{\ttfamily arXiv:1307.2199 [hep-th]}}.

\bibitem{Arkani-Hamed:2019plo}
N.~Arkani-Hamed, S.~He, T.~Lam, and H.~Thomas, ``{Binary Geometries,
  Generalized Particles and Strings, and Cluster Algebras},''
  \href{http://arxiv.org/abs/1912.11764}{{\ttfamily arXiv:1912.11764
  [hep-th]}}.

\bibitem{Arkani-Hamed:2020tuz}
N.~Arkani-Hamed, S.~He, and T.~Lam, ``{Cluster configuration spaces of finite
  type},'' \href{http://arxiv.org/abs/2005.11419}{{\ttfamily arXiv:2005.11419
  [math.AG]}}.

\bibitem{Salvatori:2018aha}
G.~Salvatori, ``{1-loop Amplitudes from the Halohedron},''
  \href{http://dx.doi.org/10.1007/JHEP12(2019)074}{{\em JHEP} {\bfseries 12}
  (2019) 074}, \href{http://arxiv.org/abs/1806.01842}{{\ttfamily
  arXiv:1806.01842 [hep-th]}}.

\bibitem{Arkani-Hamed:2019vag}
N.~Arkani-Hamed, S.~He, G.~Salvatori, and H.~Thomas, ``{Causal Diamonds,
  Cluster Polytopes and Scattering Amplitudes},''
  \href{http://arxiv.org/abs/1912.12948}{{\ttfamily arXiv:1912.12948
  [hep-th]}}.

\bibitem{Zelevinskysys}
S.~Fomin and A.~Zelevinsky, ``Y-systems and generalized associahedra,''
  \href{http://dx.doi.org/doi.org/10.4007/annals.2003.158.977}{{\em Annals of
  Mathematics} {\bfseries 158} (12, 2001) },
  \href{http://arxiv.org/abs/0111053}{{\ttfamily arXiv:0111053 [hep-th]}}.

\bibitem{Arkani-Hamed:2020cig}
N.~Arkani-Hamed, T.~Lam, and M.~Spradlin, ``{Positive configuration space},''
  \href{http://arxiv.org/abs/2003.03904}{{\ttfamily arXiv:2003.03904
  [math.CO]}}.

\bibitem{Postnikov:2005}
A.~Postnikov, ``Permutohedra, associahedra, and beyond,''
  \href{http://dx.doi.org/doi.org/10.1093/imrn/rnn153}{{\em International
  Mathematics Research Notices} {\bfseries 2009} (08, 2005) },
  \href{http://arxiv.org/abs/0507163}{{\ttfamily arXiv:0507163 [math]}}.

\bibitem{Bazier-Matte:2018rat}
V.~Bazier-Matte, G.~Douville, K.~Mousavand, H.~Thomas, and E.~Y\i ld\i~r\i m,
  ``{ABHY Associahedra and Newton polytopes of $F$-polynomials for finite type
  cluster algebras},'' \href{http://arxiv.org/abs/1808.09986}{{\ttfamily
  arXiv:1808.09986 [math.RT]}}.

\bibitem{shi1986kazhdan}
J.-Y. Shi, ``The kazhdan-lusztig cells in certain affine weyl groups,''
  \href{http://dx.doi.org/doi.org/10.1007/BFb0074968}{{\em Lecture Notes in
  Mathematics} {\bfseries 1179} (1986) XII, 312}.

\bibitem{Postnikov:2006}
A.~Postnikov, V.~Reiner, and L.~Williams, ``Faces of generalized
  permutohedra,'' {\em Documenta Mathematica} {\bfseries 13} (10, 2006) ,
  \href{http://arxiv.org/abs/0609184}{{\ttfamily arXiv:0609184 [math]}}.
  \url{http://cds.cern.ch/record/982477}.

\bibitem{Li:2018mnq}
Z.~Li and C.~Zhang, ``{Moduli Space of Paired Punctures, Cyclohedra and
  Particle Pairs on a Circle},''
  \href{http://dx.doi.org/10.1007/JHEP05(2019)029}{{\em JHEP} {\bfseries 05}
  (2019) 029},
\href{http://arxiv.org/abs/1812.10727}{{\ttfamily arXiv:1812.10727 [hep-th]}}.

\bibitem{Gao:2017dek}
X.~Gao, S.~He, and Y.~Zhang, ``{Labelled tree graphs, Feynman diagrams and disk
  integrals},'' \href{http://dx.doi.org/10.1007/JHEP11(2017)144}{{\em JHEP}
  {\bfseries 11} (2017) 144},
\href{http://arxiv.org/abs/1708.08701}{{\ttfamily arXiv:1708.08701 [hep-th]}}.

\bibitem{He:2018pue}
S.~He, G.~Yan, C.~Zhang, and Y.~Zhang, ``{Scattering Forms, Worldsheet Forms
  and Amplitudes from Subspaces},''
  \href{http://dx.doi.org/10.1007/JHEP08(2018)040}{{\em JHEP} {\bfseries 08}
  (2018) 040}, \href{http://arxiv.org/abs/1803.11302}{{\ttfamily
  arXiv:1803.11302 [hep-th]}}.

\bibitem{early2017generalized}
N.~Early, ``Generalized permutohedra, scattering amplitudes, and a cubic
  three-fold,'' \href{http://arxiv.org/abs/1709.03686}{{\ttfamily
  arXiv:1709.03686 [math.CO]}}.

\bibitem{Early:2018zuw}
N.~Early, ``{Generalized permutohedra in the kinematic space},''
\href{http://arxiv.org/abs/1804.05460}{{\ttfamily arXiv:1804.05460 [math.CO]}}.

\bibitem{He:2018okq}
S.~He and C.~Zhang, ``{Notes on Scattering Amplitudes as Differential Forms},''
  \href{http://dx.doi.org/10.1007/JHEP10(2018)054}{{\em JHEP} {\bfseries 10}
  (2018) 054},
\href{http://arxiv.org/abs/1807.11051}{{\ttfamily arXiv:1807.11051 [hep-th]}}.

\bibitem{Banerjee:2018tun}
P.~Banerjee, A.~Laddha, and P.~Raman, ``{Stokes polytopes: the positive
  geometry for $\phi^{4}$ interactions},''
  \href{http://dx.doi.org/10.1007/JHEP08(2019)067}{{\em JHEP} {\bfseries 08}
  (2019) 067}, \href{http://arxiv.org/abs/1811.05904}{{\ttfamily
  arXiv:1811.05904 [hep-th]}}.

\bibitem{Raman:2019utu}
P.~Raman, ``{The positive geometry for $\phi^{p}$ interactions},''
  \href{http://dx.doi.org/10.1007/JHEP10(2019)271}{{\em JHEP} {\bfseries 10}
  (2019) 271}, \href{http://arxiv.org/abs/1906.02985}{{\ttfamily
  arXiv:1906.02985 [hep-th]}}.

\bibitem{Chapoton}
F.~Chapoton, ``Stokes posets and serpent nests,'' \href{http://dx.doi.org/doi:
  10.1007/s10468-011-9304-4}{{\em Discrete Mathematics and Theoretical Computer
  Science} {\bfseries 18} (05, 2015) },
  \href{http://arxiv.org/abs/1505.05990}{{\ttfamily arXiv:1505.05990 [math]}}.

\bibitem{Baryshnikov}
Y.~Baryshnikov, ``On stokes sets,''
  \href{http://dx.doi.org/10.1007/978-94-010-0834-1_3}{{\em New developments in
  singularity theory’ (Cambridge, 2000), NATO Sci. Ser. II Math. Phys. Chem.}
  {\bfseries 21} (01, 2001) }.

\bibitem{saito2013grobner}
M.~Saito, B.~Sturmfels, and N.~Takayama,
  \href{http://dx.doi.org/10.1007/978-3-662-04112-3}{{\em Gr{\"o}bner
  deformations of hypergeometric differential equations}}, vol.~6.
\newblock Springer Science \& Business Media, 2013.

\bibitem{gelfand1990generalized}
I.~M. Gel'fand, M.~M. Kapranov, and A.~V. Zelevinsky, ``Generalized euler
  integrals and a-hypergeometric functions,''
  \href{http://dx.doi.org/10.1016/0001-8708(90)90048-R}{{\em Advances in
  Mathematics} {\bfseries 84} no.~2, (1990) 255--271}.

\bibitem{delaCruz:2019skx}
L.~de~la Cruz, ``{Feynman integrals as A-hypergeometric functions},''
  \href{http://dx.doi.org/10.1007/JHEP12(2019)123}{{\em JHEP} {\bfseries 12}
  (2019) 123},
\href{http://arxiv.org/abs/1907.00507}{{\ttfamily arXiv:1907.00507 [math-ph]}}.

\end{thebibliography}\endgroup
\end{document}